\documentclass[fleqn,usenatbib]{mnras}

\usepackage{newtxtext,newtxmath}
\usepackage[T1]{fontenc}

\DeclareRobustCommand{\VAN}[3]{#2}
\let\VANthebibliography\thebibliography
\def\thebibliography{\DeclareRobustCommand{\VAN}[3]{##3}\VANthebibliography}

\usepackage{graphicx}	% Including figure files
\usepackage{amsmath}	% Advanced maths commands
\usepackage{subfigure}
\usepackage{caption}
\usepackage{subcaption}
\usepackage{chngcntr} 

\usepackage{lineno}
\newcommand{\fo}{\ensuremath{f^\parallel}}
\newcommand{\fe}{\ensuremath{f^\perp}}
\newcommand{\pq}{\ensuremath{P_Q}}
\newcommand{\pu}{\ensuremath{P_U}}
\newcommand{\nq}{\ensuremath{N_Q}}
\newcommand{\nuu}{\ensuremath{N_U}}
\newcommand{\ainv}{\ensuremath{\alpha_{0}} }
\newcommand{\amin}{\ensuremath{\alpha_{\rm{min}}} }
\newcommand{\amax}{\ensuremath{\alpha_{\rm{max}}} }
\newcommand{\pa}{\ensuremath{\alpha}}
\newcommand{\pmax}{\ensuremath{P_{\rm{max}}} }
\newcommand{\pmin}{\ensuremath{P_{\rm{min}}} }

\title[Imaging Polarimetry of 67P]{Imaging Polarimetry of Comet 67\,P/Churyumov–Gerasimenko: Homogeneous Distribution of Polarisation and its Implications }

\author[Z. Gray et al.]{
Zuri Gray$^{1,2,3}$\thanks{E-mail: zuri.gray@armagh.ac.uk},
Stefano Bagnulo$^{2,1}$,
Hermann Boehnhardt$^{2}$,
Galin Borisov$^{4,2}$,
Geraint H. Jones$^{1,3}$,
\newauthor
Ludmilla Kolokolova$^{5}$,
Yuna G. Kwon$^{6}$,
Fernando Moreno$^{7}$,
Olga Mu\~{n}oz$^{7}$,
Rok Ne\v{z}i\v{c}$^{2}$,
Colin Snodgrass$^{8}$
\\
$^{1}$Mullard Space Science Laboratory, University College London, Holmbury St. Mary, Dorking RH5 6NT, UK\\
$^{2}$Armagh Observatory and Planetarium, College Hill, Armagh BT61 9DG, Northern Ireland, UK\\
$^{3}$The Centre for Planetary Sciences at UCL/Birkbeck, London WC1E 6BT, UK\\
$^{4}$Institute of Astronomy and National Astronomical Observatory, Bulgarian Academy of Sciences, 72 Tsarigradsko Chauss{\'e}e Blvd., BG-1784 Sofia, Bulgaria\\
$^{5}$Department of Astronomy, University of Maryland, College Park, MD 20742, USA\\
$^{6}$Caltech/IPAC, 1200 E California Blvd, MC 100-22, Pasadena, CA 91125, USA\\
$^{7}$Instituto de Astrofisica de Andalucia, CSIC, Glorieta de la Astronomia s/n, E-18008 Granada, Spain\\
$^{8}$Institute for Astronomy, University of Edinburgh, Royal Observatory, Edinburgh EH9 3HJ, UK
}

\date{Accepted 2024 May 13. Received 2024 May 10; in original form 2024 Janurary 18.}

\pubyear{2024}

\begin{document}
\label{firstpage}
\pagerange{\pageref{firstpage}--\pageref{lastpage}}
\maketitle

\begin{abstract}
Comet 67P/Churyumov-Gerasimenko (67P) become observable for the first time in 2021 since the Rosetta rendezvous in 2014--16. Here, we present pre-perihelion polarimetric measurements of 67P from 2021 performed with the Very Large Telescope (VLT), as well as post-perihelion polarimetric measurements from 2015--16 obtained with the VLT and the William Herschel Telescope (WHT). This new data covers a phase angle range of $\sim 4-50^{\circ}$ and presents polarimetric measurements of unprecedentedly high S/N ratio. Complementing previous measurements, the polarimetric phase curve of 67P resembles that of other Jupiter family comets and high-polarisation, dusty comets. Comparing pre- and post-perihelion data sets, we find only a marginal difference between the polarimetric phase curves. In our imaging maps, we detect various linear structures produced by the dust in the inner coma of the comet. Despite this, we find a homogeneous spread of polarisation around the photocentre throughout the coma and tail, in contrast to previous studies. Finally, we explore the consequences of image misalignments on both polarimetric maps and aperture polarimetric measurements. 
\end{abstract}

\begin{keywords}
 Comets: general -- Comets: individual: 67P/Churyumov-Gerasimenko -- Methods: observations -- Techniques: polarimetric -- Polarisation
\end{keywords}

%%%%%%%%%%%%%%%%%%%%%%%%%%%%%%%%%%%%%%%%%%%%%%%%%%

%%%%%%%%%%%%%%%%% BODY OF PAPER %%%%%%%%%%%%%%%%%%

\section{Introduction}
Comet 67P/Churyumov-Gerasimenko (hereafter 67P) is a Jupiter family comet (JFC) with an orbital period of approximately $6.45$ years, and is famously known as the target of the Rosetta mission conducted by the European Space Agency (ESA). During the two-year rendezvous, the Rosetta orbiter closely monitored the behaviour of the coma as 67P approached, passed through, and moved away from perihelion, while the lander (Philae) carried instruments for imaging and sampling the comet's nucleus. This event provided a rare opportunity for both on-site observations of the inner coma and nucleus of 67P and ground-based observations of the scattered light to be performed simultaneously \citep{snodgrass2017}. In preparation of the Rosetta mission, a world-wide observing campaign was coordinated to observe 67P, both during its 2005--2006 apparition to carefully analyse its properties prior to mission arrival at the comet, and its 2015--2016 apparition when the rendezvous took place. This campaign included most major observatories and employed all possible techniques, including polarimetry.

Polarimetric investigations exploit the fact that sunlight becomes partially polarised when it is scattered by a surface or particles. The resulting degree of polarisation depends on the physical properties of the scattering medium, including the particle size, structure, and refractive index, hence revealing clues to the global characteristics of the astronomical object. In the case of comets, light is scattered by the dust particles lifted from the nucleus by expanding gas from the sublimation of ices. The degree of linear polarisation of comets is typically measured as a function of phase angle (\pa) -- the Sun-Earth angle as seen from the comet. The polarimetric phase curves of all comets are characterised by a shallow negative polarisation branch (NPB) between $0^{\circ}$ phase angle and the so-called inversion angle \ainv (typically $\sim 22^{\circ}$), after which the polarisation is positive and increases almost linearly until reaching a maximum around phase angles $90-95^{\circ}$. The detailed morphology of the polarimetric phase curve of a comet, including the minimum and maximum values of polarisation (\pmin\ and \pmax), the phase angles these occur at (\amin\ and \amax), as well as \ainv, and the slope of the curve at this angle $h$ will vary according to the physical properties of cometary dust \citep[and references within]{kiselev2015}. Importantly, clear trends in the phase angle and wavelength dependence of polarisation can be seen among comets, which can be used to identify groups and determine some global characteristics. Except for a few notable exceptions, most comets can be categorised according to their maximum degree of polarisation \pmax. High-polarisation comets are characterised by $\pmax \sim 25-30\%$, while it varies between $\sim 8$ and $22\%$ for low-polarisation comets. Two other special classes have been specified for polarimetrically unique comets Hale-Bopp (C/1995 O1) \citep{tanga1997,kiselev1997,hadamcik2003b,hadamcik2003c}, which exhibits a polarisation degree $\sim 4\%$ higher than that for other high-polarisation comets at phase angles $34-49^{\circ}$, and comet 21P/Giacobini-Zinner, which shows low polarisation in the red domain and high polarisation in the blue \citep{kiselev2000}. Interstellar comet 2I/Borisov also exhibits a higher polarisation than typical comets, fitting into the Hale-Bopp-like comet category \citep{bagnulo2021}. %Further, according to the authors, the homogeneous spread of polarisation throughout the coma suggests that 2I/Borisov is a pristine object.

In 2021, 67P returned to the inner Solar System for the first time since the Rosetta mission, reaching perihelion on 2021-Nov-11. Here, we report our polarimetric observations of 67P during its 2015-16 and 2021 apparitions. In this latter apparition, 67P appeared brighter than in any previous apparitions and, hence, our data from this period have much higher signal-to-noise (S/N) ratio  compared to previous observations of the comet. In Sect. 2, we describe the observations and instrumentation used. In Sect. 3, we describe the methods used for polarimetric analysis. In Sect. 4, we present and discuss our findings. In Sect. 5, we conduct a qualitative study of our findings compared to previous polarimetric studies and measure the effects of image misalignments on the polarisation of 67P. In Sect. 6, we summarise and conclude this paper. 

%%%%%%%%%%%%%%%%%%%%%%%%%%%%%%%%%%%%%%%%%%%%%%%%%%

\section{Observations}

% NOTE: 2021 perihelion == 2021-Nov-02, and 2015 perihelion == 2015-Aug-13
Our observations were performed using two instruments: FORS2 (FOcal Reducer/low dispersion Spectrograph 2) at the VLT (Very Large Telescope) of ESO (European Southern Observatory), and ISIS (Intermediate-dispersion Spectrograph and Imaging System) at the WHT (William Herschel Telescope) of the ING (Isaac Newton Group of Telescopes). Our observations are summarised in Table \ref{tab:obslog}.

The polarimetric set-up of FORS2 follows the dual-beam polarimeter design first proposed by \cite{appenzeller1967}. A rotatable retarder wave plate, $\lambda/2$ for the case of linear polarisation, is used to introduce a phase shift between the electric field components aligned parallel and perpendicular to the optical axis. A Wollaston prism is used as a beam splitting device, separating the incoming light into two orthogonally polarised beams: the ordinary and extraordinary beams, for which we will adopt the notation \fo\ and \fe. A Wollaston mask, consisting of nine $6.8' \times 22''$ strips, is introduced to prevent the superposition of the \fo\ and \fe\ beams. The field of view (FoV) ($6.8' \times 6.8'$ in regular imaging mode) is vignetted as a result, which can be inconvenient in the case of extended objects such as comets as the outer edges of the coma can be cutout from view. Finally, the detector is a 2k $\times$ 4k CCD with a pixel scale of $0.125" \rm{pixel}^{-1}$ for $1 \times 1$ binning, and $0.25" \rm{pixel}^{-1}$ for $2 \times 2$ binning. This particular setup allows the user to implement the beam-swapping technique \citep{bagnulo2009}, which largely suppresses the effects of instrumental polarisation. For this, we took observations with the retarder waveplate at eight different angles between $0-157.5^{\circ}$, increasing in steps of $22.5^{\circ}$. Throughout our observations, we used various filters: broadband filters centred around 557\,nm, 655\,nm, and 768\,nm, known as v\_HIGH, R\_SPECIAL,and I\_BESS filters, respectively (V-, R- and I-filter hereafter), as well as non-standard filters centred around 485\,nm and 834\,nm, known as FILT\_485\_37 and FILT\_834\_48 (B- and IR-filter hereafter). Differential tracking was used during observations to account for the motion of the comet, ensuring that the photocentre of the comet consistently occupied the same few pixels (both \fo\ and \fe\ beams) in the centre of the CCD across all observations. As described in \cite{patat2006} and \cite{gonzalez2020},  the FORS2 instrumental polarisation remains minimal with minimal variations in the central region of the CCD across various filters. The observing conditions were clear with sub-1 arcsec seeing for almost all observations, with the exception of 2021-Jul-18 which had some thin clouds and $\sim1.1-1.3$ arcsec seeing at the time of the observations, and 2021-Aug-05 which had $\sim1.3-1.5$ arcsec seeing. 
%In imaging polarimetric mode, acquisition images are also taken free of polarimetric optics before they are mounted for the observations. Therefore, we have also acquired an image without the vignetting of the Wollaston mask for each observing epoch. Although these images have low S/N ratio due to the short exposure time (one minute or less), they are useful for analysis of the coma. 

ISIS is a highly efficient, double-arm, low to mid-resolution spectrograph with a long-slit observing window of up to $\sim 4'$ in length, and $\sim 22'$ in width. It is converted into a spectro-polarimeter when equipped with additional polarimetric optics, following a scheme conceptually similar to that of FORS2 but using a Savart plate instead of a Wollaston prism. The insertion of a mirror on the top of one of the grisms transforms the instrument into a imaging-polarimeter with a pixel scale of $0.22 "\rm{pixel}^{-1}$. Our polarimetric observations were carried out in a similar fashion to that of FORS2, with the retarder waveplate at eight angles between $0^{\circ} - 157.5^{\circ}$ and with the Sloan Gunn R filter (which we will also refer to as R-filter) in the red arm. Finally, we observed high polarised standard stars on each of the nights of the ISIS observations to account for instrumental polarisation. These observations are summarised in Table \ref{tab:starobslog}.

%%%%%%%%%%%%%%%%%%%%%%%%%%%%%%%%%%%%%%%%%%%%%%%%%%

%%%%%%%%%%%%%%%%%%%%%%%%%%%%%%%%%%%%%%%%%%%%%%%%%%

\section{Data Reduction}

\subsection{Polarimetry: Definitions}
To measure the degree of linear polarisation, we employed the so-called beam-swapping technique, as suggested explicitly in the FORS2 user manual and discussed thoroughly by \cite{bagnulo2009}. Following this method, we measured the reduced Stokes parameters, $\pq' = Q/I$ and $\pu' = U/I$ (where $Q$ and $U$ are the Stokes parameters as defined in \cite{shurcliff1962}), by combining a series of exposures taken with the retarder waveplate rotated at different position angles using the following equations:
\begin{equation}
\label{eq:pqpu}
\begin{split}
    \pq' &= \frac{1}{2} [D (\phi = 0^{\circ}) + D (\phi = 90^{\circ})], \\
    \pu' &= \frac{1}{2} [D (\phi = 22.5^{\circ}) + D (\phi = 112.5^{\circ})],
\end{split}
\end{equation}
and 
\begin{equation}
\label{eq:D}
D (\phi) = \frac{1}{2} \left[ \left( \frac{f^{\parallel} - f^{\perp}}{f^{\parallel} + f^{\perp}} \right) _{\phi} - \left( \frac{f^{\parallel} - f^{\perp}}{f^{\parallel} + f^{\perp}}\right)_{\phi+45\degr} \right].
\end{equation}
Here, \fe\ and \fo\ are the photon counts of the source in the extraordinary and ordinary beams (corrected for sky background), and $\phi$ is the position angle $(0,22.5,45^{\circ}$, etc.) of the rotatable retarder waveplate. In this initial step, $\pq'$ and $\pu'$ are measured in the reference system of the instrument. To transform these parameters into the reference system of the comet, i.e., the direction perpendicular to the scattering plane, we used the equations:
\begin{equation}
\begin{split}
\pq &=\phantom{-} \cos(2\chi)\pq' + \sin(2\chi)\pu'  \\
\pu &= -\sin(2\chi)\pq' + \cos(2\chi)\pu' 
\end{split}
\label{eq:trans}
\end{equation}
where  $\chi = {\rm PA} + \Phi + 90^{\circ} + \varepsilon$\ , $\Phi$\ is the angle between the comet-North Celestial Pole direction and the comet-Sun direction, and $\varepsilon$ is a small value to correct for the slight chromatism (wavelength dependence) of the retarder waveplate. The position angle of the retarder waveplate $\phi$ indicates the orientation of the fast axis, i.e., the direction along which the electric field propagates without a phase shift. This direction is slightly wavelength dependent, which results in a minor rotation of the measured polarisation position angle. As given in the user manuals of each instrument, the $\varepsilon$ values to correct for this chromatism are $3.47, 1.80, -1.19, -2.89, -2.67$ for the FORS2 B-, V-, R- ,I-, and IR-filters, respectively, and $1.86$ for ISIS R-filter. 

After this transformation, all polarisation is expected to be in the direction perpendicular to the scattering plane. In other words, all polarisation should be in $\pq$, while $\pu$\ is expected to be zero within uncertainties. Thus, the proximity of $\pu$\ to zero can be used as a quality check for our measurements---this will be further discussed in Sect. \ref{sec_quality}. Alternatively, the polarisation can be described with the expressions:
\begin{equation}
P_L = \sqrt{\pq^{2} + \pu^{2}} \\
\end{equation}
\begin{equation}
\theta = \frac{1}{2} \rm{arctan} \frac{\pu}{\pq} + \Delta\theta
\end{equation}
where $P_L$ is the fraction of linear polarisation, not affected by the chromatism of the retarder waveplate (i.e., $P_L = P'_L$), and $\theta$ is its position angle. $\Delta\theta$ is to account for the nature of the arctan function being defined between $-90^{\circ}$ and $90^{\circ}$. Thus, $\Delta\theta = 0$ when $\pq\ > 0$, $\pi/2$ when $\pq\ < 0$, $\pi/4$ when $\pq=0$ and $\pu > 0$, and $-\pi/4$ when $\pq=0$ and $\pu < 0$. We calculated the uncertainties of the expressions defined above following standard error propagation methods discussed, for instance, in \cite{bagnulo2009} and \cite{bagnulo2017}.

%%%%%%%%%%%%%%%%%%%%%%%%%%%%%%%%%%%%%%%%%%%%%%%%%%

\subsection{Pre-processing}
All data reduction processes were performed using customised Python scripts. In preparation for subsequent analysis, we applied an initial set of image pre-processing steps to our science images. All frames were first bias-subtracted using a master bias image created for each observing night. They were then divided by a master flat-field image, created from the median of a series of exposures of a uniform source of illumination. In the case of FORS2, these were obtained on sky at twilight without polarimetric optics. For ISIS, they were obtained with the retarder waveplate continuously rotating at a frequency of 60 rpm to depolarise the sky and achieve uniform illumination of the two beams separated by the Savart plate. For the 2016-Feb-04 data set, dome-flats were obtained instead of sky-flats. 

The first step in our data reduction process was to find an accurate pixel position of the comet nucleus in both the \fe\ and \fo\ strips on all CCD images. Although the comet nucleus is not directly observable, it is reasonable to assume that it is buried within the coma photocentre. To find the position of the photocentre, we used the DAOStarFinder Python package \citep[based on ][]{stetson1987}, which searches for the local density maxima that has a peak amplitude greater than a given threshold and a size and shape similar to a defined 2D Gaussian kernel. The convoluted result is close to a Gaussian and can be used to find the centre position. We assessed the FWHM values of several stars across the CCD to determine suitable values to use in our algorithm. To account for any cometary activity, which may appear as flux variations in or close to the photocentre and ultimately lead to inaccurate pixel positions, we opted for using larger full width half maximum (FWHM) values.

%%%%%%%%%%%%%%%%%%%%%%%%%%%%%%%%%%%%%%%%%%%%%%%%%%

\subsection{Aperture Polarimetry} \label{sec_AperturePol}

To calculate the reduced Stokes parameters, we measured the flux of the comet in the \fe\ and \fo\ beams within apertures of various sizes centred around the comet photocentre. 
The background sky level was measured in a detached rectangular aperture at a location free from contamination of the coma, tail and background stars, but within the central $\sim4$ arcmin of the CCD FOV, which bears minimal instrumental polarisation and minimal variations in instrumental polarisation as discussed in \cite{patat2006} and \cite{gonzalez2020}. We then combined the \fe\ and \fo\ fluxes, background subtracted, of each position of the waveplate ($0-157.5^{\circ}$) using Eq. \ref{eq:pqpu} to calculate $\pq'$ and $\pu'$. Finally, we transformed these values into the reference system of the comet using Eq. \ref{eq:trans}.

%%%%%%%%%%%%%%%%%%%%%%%%%%%%%%%%%%%%%%%%%%%%%%%%%%

\subsection{Imaging Polarimetry} \label{sec_Maps}
We created both imaging and polarimetric maps with our VLT data. We did not produce similar maps for our WHT data due to the considerably lower S/N ratio and smaller FoV of the ISIS observations compared to those of FORS2. Polarimetric maps, as well as aperture polarimetry of the whole comet, provide insight on the bulk physical properties of the dust particles comprising the coma and tail. They allow us to explore the distribution of polarisation at different photocentric distances, which may be indicative of changes in dust grain properties across the coma. As summarised in \cite{kolokolova2007}, the spatial distribution of polarisation is a useful tool to classify comets---dust rich tend to show a uniform spread of polarisation across the coma, while gas rich comets are typically characterised by a region of high polarisation close to the photocentre and a strong radial fall in polarisation. Further, as explored in \cite{hadamcik2003c}, areas of localised cometary activity sometimes show an increased level of polarisation compared to the surrounding coma. 

To create maps from our polarimetric data, we considered the \fe\ and \fo\ beams split by the Wollaston prism as two individual images. From aperture polarimetry, we already know the photometric centres of the comet and the mean background sky count in each strip of each image. With this information, we shifted the images of all strips such that the photocentres of the comet were in the exact same pixel position, as well as subtracted the background sky corresponding to each strip. This resulted in sixteen frames, corrected for background sky, aligned and ready to be combined. To generate polarimetric maps, we combined the frames with the relevant retarder waveplate position angles with Eq. \ref{eq:pqpu}, then transformed these maps into the reference direction of the comet with Eq. \ref{eq:trans}, resulting in \pq\ and \pu\ maps. For the imaging maps, we simply stacked (i.e. co-added) the sixteen comet-centred, background subtracted frames, resulting in high S/N images. 

To assess the presence of localized cometary activity, we applied two image enhancement methods to our imaging maps: (i) adaptive Laplace filtering \citep{boehnhardt1994} and (ii) coma renormalisation \citep{ahearn1986}. For this, we used the ESO-MIDAS image processing tool \citep{midas2013} The Laplacian filter enhances regions of sharp discontinuities and is very useful for detecting the presence of spatial features from local regionally enhanced activity on the nucleus of the comet. We applied multiple Laplacian filter widths in order to search for coma structures of different extensions.  Coma renormalisation mostly highlights radial features that rapidly fade with distance and the results (geometry, strength) are very sensitive to accurate centering, at least close to the coma center. On the one hand, the Laplace filtering method is very sensitive to low-level systematic brightness differences in the coma and tail, but reduces the overall spatial resolution of the given structures. On the other hand, the renormalisation method is less sensitive to low-level structures, but keeps the linear intensity scaling of the original image and does not reduce the geometric resolution. Inspections of these enhanced images are useful for assessing the variability of global coma structures. To reduce the probability of creating spurious features in the enhanced images due to the misalignment of polarimetric frames, we applied the digital features directly to the imaging maps of the comet (i.e. after combination).  

Finally, to highlight any possible low-contrast structures in the polarimetric maps, we have applied a multidimensional Gaussian filter to our polarimetric maps, essentially convolving the image with a Gaussian kernel. This process results in a smoothing effect on the image, reducing noise while preserving the overall structure and edges. To reduce the probability of creating spurious features due to the misalignment of polarimetric frames, we applied the Gaussian filter to the final polarimetric maps (i.e. after combination).

%%%%%%%%%%%%%%%%%%%%%%%%%%%%%%%%%%%%%%%%%%%%%%%%%%

\subsection{Quality Checks} \label{sec_quality}
Measuring and analysing the polarisation of comets is not a straightforward task, since (i) measurements can be prone to artefacts introduced by instrument characteristics, and (ii) less than optimal alignment of the comet photocentre when recombining various frames can create artificial artefacts. Here, we discuss the effect these issues can have and how we have handled them. 

(i) Due to the effects of instrumental polarisation, the polarised signal arriving at the CCD can be different from that incident on the telescope. We accounted for this by measuring the polarisation of highly polarised standard stars with ISIS at the time of the comet observations. We then used the difference between the measured polarisation of the standard stars to that of literature values to calibrate the instrument. This step is not necessary with FORS2, as observations of standard stars are already a part of the calibration plan of the instrument. Further, to verify the validity of the measured \pq\, both numerical values from aperture polarimetry and the polarimetric maps, we calculated the \pu\ values and maps, as well as those corresponding to the null parameters, \nq\ and \nuu, defined as:
\begin{equation}
\begin{split}
    \nq &= \frac{1}{2} [D (\phi = 0^{\circ}) - D (\phi = 90^{\circ})], \\
    \nuu &= \frac{1}{2} [D (\phi = 22.5^{\circ}) - D (\phi = 112.5^{\circ})].
\end{split}
\label{eq:null}
\end{equation}
Here, the quantity $D$ has the same definition as in Eq. \ref{eq:D}. \pu\ is expected to be zero due to symmetry reasons and the null parameters should also have a Gaussian distribution centred around zero with the same FWHM as their corresponding Stokes parameter. A significant deviation from zero would flag an issue with either the data (e.g. a cosmic ray) or the data reduction process. Since we expect \pu, \nq\ and \nuu\ (both measured within an aperture and as maps) to all oscillate around zero within uncertainties, we can use these three quantities as a quality check of our measurements. 

(ii) As mentioned, we used a Python algorithm to find the position of the comet photocentre in the \fe\ and \fo\ strips of each frame. The point spread function (PSF) of the photocentre as a whole is the sum of the nucleus (a point-source), the coma profile, as well as any flux variations attributed to intrinsic structures of the comet (jets, fans, etc.)---this latter point may cause issues when calculating an accurate pixel position of the photocentre. Inaccurate calculations of the pixel positions of the photocentre will ultimately lead to misalignments when combining frames to create imaging and polarisation maps. Consequently, spurious features can be generated close to the photocentre, which may be mistaken for true features intrinsic to the cometary coma. In Sect. \ref{sec_misalignment}, we investigate the consequences of deliberately introducing random minor misalignments, ranging between (--1,1) pixels in the x- and y-directions, to the position of the comet photocentre.

%%%%%%%%%%%%%%%%%%%%%%%%%%%%%%%%%%%%%%%%%%%%%%%%%%
%%%%%%%%%%%%%%%%%%%%%%%%%%%%%%%%%%%%%%%%%%%%%%%%%%

\section{Results and Discussion}

\subsection{Aperture Polarimetry}

\begin{figure*}
\begin{center}
    \centering
    \includegraphics[width=\textwidth]{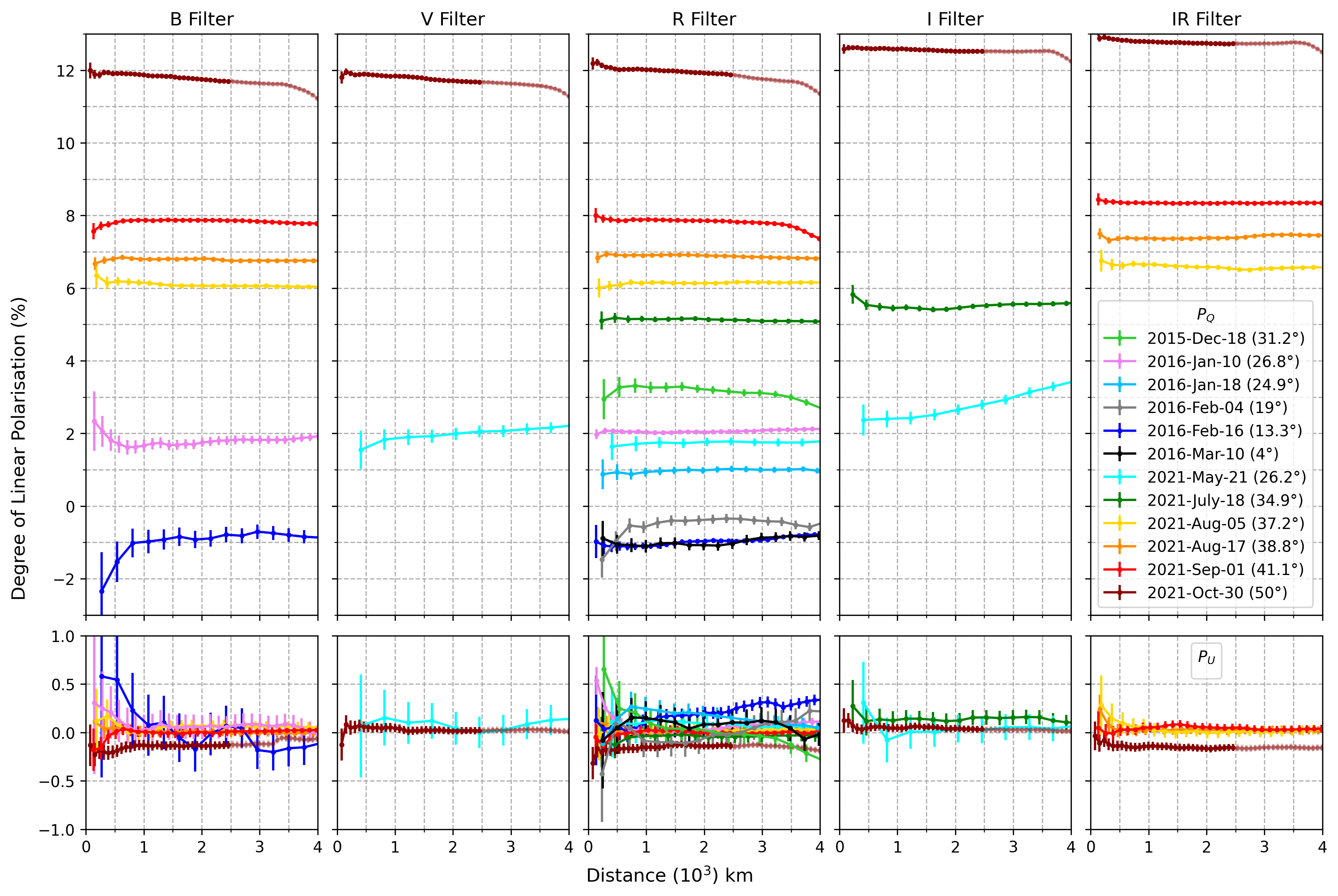}
    \caption{Aperture polarimetric measurements of all 67P data, i.e. the polarisation measured as a function of aperture radius, increasing in increments of one pixel. In cases where error bars are not shown, they are smaller than the data point. The top panels present the \pq\ measurements, while the lower panels present \pu\ (both rotated to be perpendicular to the scattering plane). All measurements have been scaled to the same distance scale (horizontal axis) and the values reported in Table \ref{tab:obslog} are those integrated in an aperture radius equivalent to 2000 km. For FORS2, the R, B, IR, I, and V filters refer to the R\_SPECIAL, FILT\_485\_37, FILT\_834\_48, I\_BESS, and V\_BESS respectively, and for ISIS, the R refers to Sloan Gunn R filter. A number of measurements are somewhat spurious, as described in the text.}
    \label{fig:growthcurves}
\end{center}
\end{figure*}

\begin{table*}
\begin{center}
\caption{\label{tab:obslog}
Observing log of comet 67P. $\Delta$ is the apparent range of the target center relative to the Earth, the scale is the kilometers covered by $1"$ at each $\Delta$ value (note: the pixel scale of FORS2 and ISIS are $0.25" \rm{pixel}^{-1}$ and $0.22" \rm{pixel}^{-1}$, respectively), r is the Sun's apparent range relative to the target center as seen from the Earth, $\alpha$ is the phase angle, INSTR and FILT indicate the instrument and filter used for a given observation, EXP is the exposure time of each polarimetric frame; and \pq, \pu, $P_L$, and $\theta$ are the polarimetric measurements indicated in the text. For FORS2, the R, B, IR, I, and V filters refer to the R\_SPECIAL, FILT\_485\_37, FILT\_834\_48, I\_BESS, and V\_BESS respectively, and for ISIS, the R refers to Sloan Gunn R filter.
}
\begin{small}
  \begin{tabular}{rcccccllcr@{$\pm$}lr@{$\pm$}lr@{$\pm$}lr }
    \hline\hline
    DATE & Days from & $\Delta$ & Scale & r & $\alpha$ & INSTR & FILT & EXP & \multicolumn{2}{c}{\pq} & \multicolumn{2}{c}{\pu} & \multicolumn{2}{c}{$P_L$} & $\theta$ \\
     & Perihelion & (AU) & (km/") & (AU) & $(^{\circ})$ & & & (s) & \multicolumn{2}{c}{(\%)} & \multicolumn{2}{c}{(\%)} & \multicolumn{2}{c}{(\%)} & $(^{\circ})$ \\ 
    \hline
    2015-Dec-18 & +127 & 1.69 & 122 & 1.9  & 30.9 & ISIS  & R & 120 &3.19 & 0.11  & 0.01  & 0.11  & 3.19  & 0.11  & 0.09 \\
    2016-Jan-10 & +150 & 1.58 & 115 & 2.09 & 26.8 & FORS2 & B & 300 & 1.75  & 0.13  & 0.05  & 0.12  & 1.75  & 0.13  & 0.86 \\
               &      &      &      &      &&       & R & 200 & 2.04  & 0.05  & 0.02  & 0.05  & 2.04  & 0.05  & 0.34 \\ [2mm] 

    2016-Jan-18 & +158 & 1.55 & 112 & 2.14 & 24.6 & ISIS & R & 300 &$0.98 $& 0.09  & 0.19  & 0.09  & 1.0   & 0.09  & 5.49 \\
    2016-Feb-04 & +175 & 1.5 & 108 & 2.27  & 18.5    & ISIS & R & 300 &$-0.39$& 0.13  & -0.08 & 0.14  & 0.4   & 0.13  & 5.80 \\
    2016-Feb-16 & +187 & 1.49 & 108 & 2.37 & 13.3  & FORS2& B & 300 &$-0.87$  & 0.18  & $-0.11$  & 0.18  & 0.88  & 0.18  & 3.7 \\
               &      &      &      &       &&      & R & 650 &$-0.96$  & 0.10  & 0.2  & 0.10  & 0.98  & 0.10  & $-5.93$ \\[2mm]

    2016-Mar-11 & +210 & 1.56 & 113 & 2.55 & 3.8     & ISIS & R & 240 &$-1.08$& 0.14  & 0.08  & 0.15  & 1.08& 0.14  &177.78\\
\hline
    2021-May-21 &$-177$& 2.26 & 164 & 2.21 & 26.15 & FORS2 & V &  80 & 1.99  & 0.17  & 0.04  & 0.17  & 1.99  & 0.17  & 0.63  \\ 
               &      &      &      &       & &      & R & 200 & 1.77  & 0.12  & 0.09  & 0.12  & 1.77  & 0.12  & 1.43  \\
               &      &      &      &       &   &    & I & 100 & 2.65  & 0.14  & 0.04  & 0.14  & 2.65  & 0.14  & 0.44\\[2mm] 

    2021-Jul-18 &$-119 $& 1.27& 92 & 1.76 & 34.9  & FORS2 & R &  60 & 5.14  & 0.06  & $-0.05$  & 0.07  & 5.14  & 0.06  & $-0.27$  \\
               &      &      &      &       &     &  & I &  50 & 5.46  & 0.07  & 0.12  & 0.08  & 5.46  & 0.07  & 0.65  \\[2mm]

    2021-Aug-05 &$-101$& 1.01 & 73 & 1.61 & 37.23 & FORS2 & B & 120 & 6.07  & 0.06  & 0.07  & 0.06  & 6.07  & 0.06  & 0.32  \\ 
               &      &      &      &       & &      & R &  20 & 6.14  & 0.05  & 0.02  & 0.05  & 6.14  & 0.05  & 0.11  \\ 
               &      &      &      &       &   &    & IR&  70 & 6.58  & 0.06  & -0.01  & 0.06  & 6.58  & 0.06  & -0.02\\[2mm]

    2021-Aug-17 &$-89 $& 0.87 & 63 & 1.53 & 38.83 & FORS2 & B & 120 & 6.81  & 0.05  & 0.01  & 0.04  & 6.81  & 0.05  & 0.04 \\
               &      &      &      &       &     &  & R &  20 & 6.89  & 0.04  & 0.02  & 0.04  & 6.89  & 0.04  & 0.07  \\ 
               &      &      &      &       &       && IR&  70 & 7.38  & 0.04  & 0.03  & 0.04  & 7.38  & 0.04  & 0.11 \\ [2mm]

    2021-Sep-01 &$-74 $& 0.71 & 51 & 1.43 & 41.1  & FORS2 & B &  80 & 7.87  & 0.04  & 0.01  & 0.04  & 7.87  & 0.04  & 0.04 \\
               &      &      &      &       & &      & R &  10 & 7.86  & 0.04  & $-0.01$  & 0.04  & 7.86  & 0.04  & $-0.01$ \\ 
               &      &      &      &       &  &     & IR&  50 & 8.34  & 0.04  & 0.05  & 0.04  & 8.34  & 0.04  & 0.18 \\ [2mm]

    2021-Oct-30 &$-15 $& 0.42 & 31 & 1.21 & 50    & FORS2 & B &  15 &11.76  & 0.04  & -0.14  & 0.04  & 11.76  & 0.04  & -0.33 \\ 
               &      &      &      &       &  &     & V &   5 &11.71  & 0.04  & 0.02  & 0.04  & 11.71  & 0.04  & 0.05 \\
               &      &      &      &       &   &    & R &   3 &11.92  & 0.04  & -0.14  & 0.04  & 11.93  & 0.04  & -0.33 \\ 
               &      &      &      &       &    &   & I &   5 &12.53  & 0.04  & 0.04  & 0.04  & 12.53  & 0.04  & 0.1 \\ 
               &      &      &      &       &     &  & IR&  13 & 12.74  & 0.04  & -0.16  & 0.04  & 12.75  & 0.04  & -0.37 \\ 
        \hline
  \end{tabular}

\end{small}
\end{center}
\end{table*}

\begin{table*}
\begin{center}
  \caption{\label{tab:starobslog} Polarimetric observations of standard stars with ISIS. The literature values of NGC 2041-1 were taken from \protect\cite{fossati2007} and values of HD 251204 from \protect\cite{weitenbeck1999}.}
  \centering
  \begin{tabular}{crcr@{$\pm$}lr@{$\pm$}lr@{$\pm$}lr@{$\pm$}l}
  \hline \hline
  & & & \multicolumn{4}{c}{LITERATURE} & \multicolumn{4}{c}{THIS WORK} \\
  STAR & DATE & FILTER & \multicolumn{2}{c}{$P_L$ ($\%$)} & \multicolumn{2}{c}{$\Theta (^{\circ})$} & \multicolumn{2}{c}{$P_L$ ($\%$)} & \multicolumn{2}{c}{$\Theta (^{\circ})$}  \\
  \hline
    NGC 2041-1 & 2015-Dec-18 &  R & $9.62$ &  $0.01$ & $135.84$ &  $40.02$ & $9.89$ &  $0.04$ & $138.68$ &  $0.1$ \\
               & 2016-Jan-18 &  R &\multicolumn{4}{c}{}                   & $9.89$ &  $0.04$ & $138.16$ &  $0.1 $ \\
               & 2016-Mar-10 &  R &\multicolumn{4}{c}{}              & $9.7$ &  $0.05$ & $138.57$ &  $0.1 $ \\ [2mm]

    HD 251204  & 2016-Jan-18 &  R & $4.72$ &  $0.04$ & $153.3$ &  $0.26$  & $4.78$ &  $0.02$ & $156.86$ &  $0.1 $ \\
               & 2016-Mar-10 &  R &\multicolumn{4}{c}{}                & $4.92$ &  $0.05$ & $156.42$ &  $0.1 $ \\
    \hline
    \end{tabular}
\end{center}
\end{table*}

In Fig. \ref{fig:growthcurves}, we plot the aperture polarimetric measurements---\pq\ (upper panel) and \pu\ (lower panel) measured as a function of aperture radius increasing in steps of one pixel, as well as the error calculated from photon noise and background subtraction. Considering the varying comet-Earth distance ($\Delta$) throughout the observing campaign, we have scaled all measurements to the same distance on the horizontal axis. These measurements show that \pq\ sometimes depends on the aperture size, while \pu\ indicates the reliability of these measurements. For point sources, variations in \pq\ with increasing aperture are likely due to the presence of background stars whose flux is included within the aperture measurement. In the case of active comets, however, it is reasonable to presume that such variations are intrinsic to the comet itself, reflecting variations in the dust properties of the coma, and hence, properties of the scattered light. Despite this, the majority of our measurements show a nearly flat, smooth spread of polarisation around the photocentre with variations within each case rarely exceeding $\sim 0.5\%$. 

This is not true for a few cases, which we explain by the following. The increase in \pq\ with aperture size in the I-filter 2021-May-21 epoch is due to the proximity of a bright background star, whose flux contaminates that of the comet. The reason this feature does not appear in the May epochs of the other filters is because observations in various filters are taken sequentially rather than simultaneously. The drop in \pq\ at larger aperture sizes across all filters of 2021-Oct-30 is due to the inclusion of the edge of the Wollaston mask in the aperture---this may be more clear upon inspecting the imaging and polarimetric maps in Figs. \ref{fig:IntMaps} and \ref{fig:PolQMap}. For this reason, we do not consider values above an aperture radius of $\sim 2500$ km, which we indicate in Fig. \ref{fig:growthcurves} with different shading. 

In addition, we explain the deviations in \pu\ from zero with the following. Both B- and R-filter observations of the 2016-Feb-16 were contaminated by a bright background star, leading to non-zero \pu\ values. The systematically larger \pu\ values measured in the B-, R-, and IR-filters of 2021-Oct-30 are most likely due to a non-perfect calibration of the retarder waveplate chromatism, which may propagate into an uncertainty in \pq\ and \pu\ larger than that due to photon-noise. An alternative explanation is the imperfect alignment of the retarder waveplate, as seen in \cite{bagnulo2017}. Although not listed, we also verified that the numerical \nq\ and \nuu\ values for each corresponding Stokes measurement are close to zero within uncertainty.

In Table \ref{tab:obslog}, we record our numerical results of the linear polarisation of 67P, measured within a circular aperture corresponding to a 2000\,km radius centred around the photocentre. In Table \ref{tab:starobslog}, we record our polarimetric measurements of standard stars taken with the ISIS instrument. We found an average offset of $\theta \sim 3.04^{\circ}$ between our measurements and those found in the literature, which we used this value to calibrate our data. %However, this calibration only corrects the instrumental polarisation in the central region of the CCD where the target is located during observations. \cite{patat2006} have shown that FORS2 suffers from an instrumental polarisation that increases with distances from the centre of the CCD. For extended objects like comets, this causes an issue since the background sky level must be measured in a location free from the contamination of the comet and tail. In some cases, this means measuring the background several hundred pixels away from the CCD centre. Therefore, taking into account both these factors, we were careful to select the optimal position to measure the background sky in each frame.

Gas molecules present in the cometary atmosphere can partially depolarise the light scattered by the dust component \citep{kiselev2004, jewitt2004, jockers2005}. According to spectroscopic studies by \cite{ivanova2017} and \cite{opitom2017}, 67P is characterised by a strong CN emission, as well as relatively weak $\rm{C_2}$, $\rm{C_3}$, $\rm{CO^{+}}$, and $\rm{NH_2}$ emissions at certain heliocentric distances. The CN emission, however, falls outside the wavelength range of the filters used for our observation. On the other hand, the $\rm{C_2}$, $\rm{C_3}$, $\rm{CO^{+}}$, and $\rm{NH_4}$ emissions fall within the wavelength bands of the B, V, and R-filters. Considering their weak emissions, we can assume their depolarising effect is very small or negligible in most cases, though may increase as comet activity increases on approach to perihelion. 

%%%%%%%%%%%%%%%%%%%%%%%%%%%%%%%%%%%%%%%%%%%%%%%%%%

\subsection{Phase Angle Dependence of Polarisation}
\begin{figure*}
%\begin{center}
    \centering
    \includegraphics[width=\textwidth]{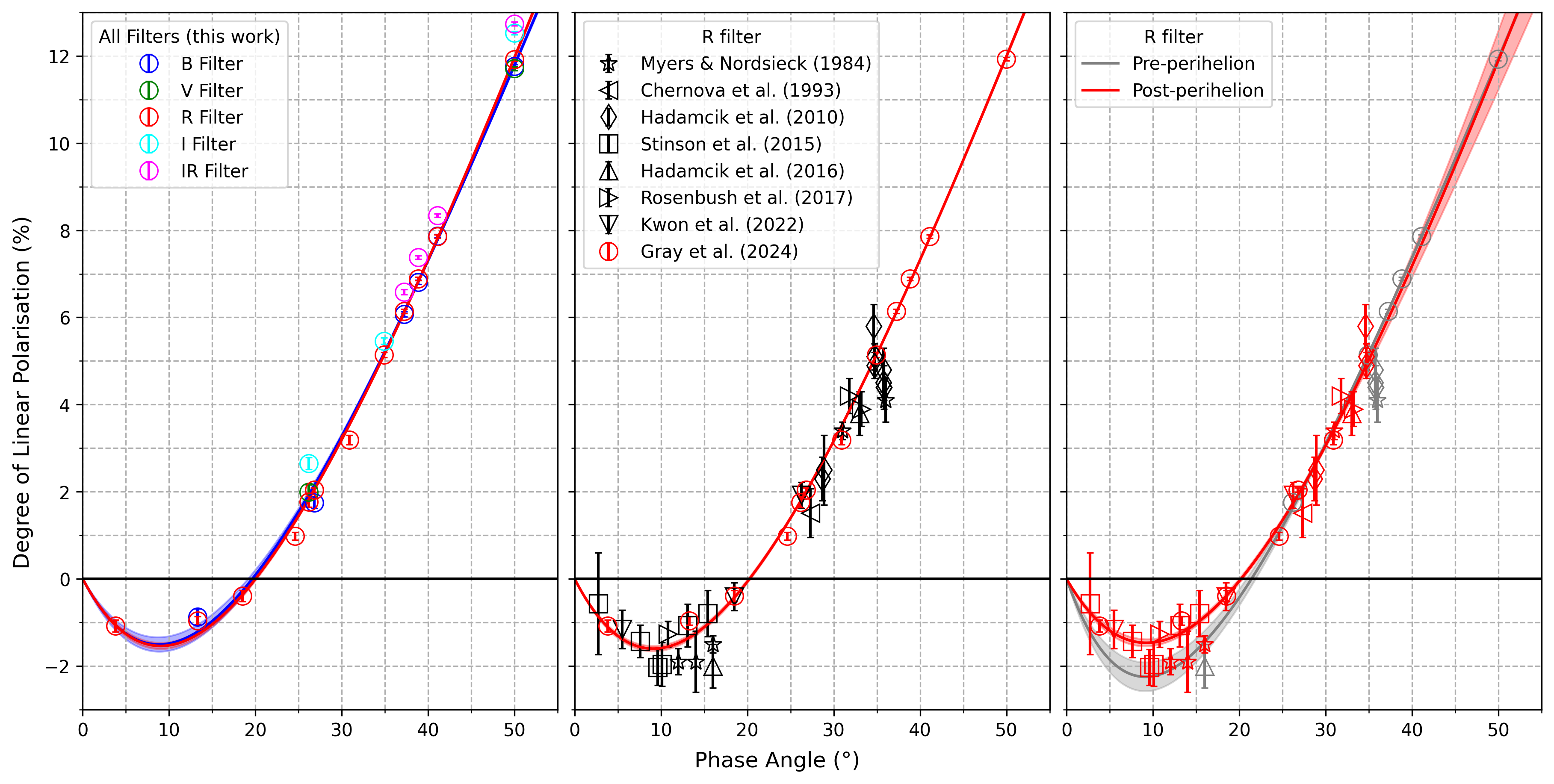}
    \caption{Polarisation phase curve of 67P. \textit{Left panel:} our measurements in all filters plotted as a function of phase angle, and their best fit. \textit{Centre panel:} our measurements in the R filter plotted with all other values available in literature. The curve is the best fit of all data in the plot. \textit{Right panel:} the same data as the centre panel, but the best fits of the pre- (red) and post-perihelion (grey) data have been calculated separately. The shaded area around the curves, where applicable, represent 1-sigma error of the best fit.}
    \label{fig:phasecurve}
%\end{center}
\end{figure*}

In Fig. \ref{fig:phasecurve}, we plot the polarimetric phase curve of 67P in three ways. The left panel displays all new \pq\ measurements presented in this work, where observations in the various filters, as well as their best fits (where applicable), are distinguished with colours. As expected, the comet shows a typical phase angle dependence of linear polarisation: the polarisation is negative at small phase angles, forming a shallow negative polarisation branch, followed by a bell-shaped positive polarisation branch. The centre panel shows our R-filter polarisation measurements (highlighted in red) compared to other R-filter measurements available in the literature (in black). In the right panel, we have separated all data into two groups, pre- and post-perihelion, and investigate their differences.

To calculate the best fits, we use the linear-exponential empirical model suggested by \cite{muinonen2009}:
\begin{equation}
    P(\alpha) = A(e^{\frac{-\alpha}{B}} - 1) + C\alpha,
    \label{eq:cellinofunc}
\end{equation}
where $\alpha$ is the phase angle, and $A, B, C$ are free parameters which shape the curve and are derived by least-square fitting. The shaded area around the curves (where applicable) represent the $1 \sigma$ error of the fit. 

% Previous perihelion dates:
% 2021-11-02
% 2015-08-13
% 2009-02-28
% 1996-01-08
% 1989-06-18
% 1982-11-03

% Aperture sizes used by other authors for comparing in this study:
% Kwon et al. (2021) ~ 1 000 km in radius
% Stinson et al. (2016) ~ 10 000 km in diameter?
% Rosenbush et al. (2017) ~  2 500 km in radius, found a variation with aperture
% Hadamcik et al. (2016) ~ 1 200 km in diameter for HST (~-2 percent). ~ 18 000 km diameter == 5.2 percent, 10 000 km diameter (3.8 percent  
% Hadamcik et al. (2010) ~ maps. pol was ~6percent in central 3000 km aperture and surround coma was around 2percent. pol decreases with radial distance. "As the particles from the jets moved away from the nucleus, they mixed with background particles with potentially different physical properties. Selection by size or some fragmentation may also occur: as the dust particles leave the inner coma, the largest particles could stay closer to the nucleus for a longer time."
% Myers & Nordsieck 1984 ~ ? (spectropol)
% Chernova

\subsubsection{Comparison with Previous Measurements of 67P}

\cite{myers1984} and \cite{chernova1993} were the first to carry out polarimetric measurements of 67P during its 1982-83 apparition. \cite{stinson2016} and \cite{hadamcik2010} measured the comet's polarisation in its 2008-2009 apparition, and finally \cite{hadamcik2016}, \cite{rosenbush2017} and \cite{kwon2022} measured it in the 2015-16 apparition. These measurements are shown in the centre panel of Fig \ref{fig:phasecurve}, alongside our new measurements (R-filter only). Here, the curve is the best fit of all the data in this plot. In the case where other studies measured the polarisation in multiple aperture sizes, we opted for choosing values measured in larger aperture sizes for our comparison---we elaborate on the reason for this choice in Sect. \ref{sec_misalignment}. In comparison to these previous measurements, our new observations feature unprecedentedly high S/N, and thus, are the most precise measurements to date. According to the best fit and $1 \sigma$ uncertainty, this polarimetric phase curve is characterised by $\pmin = (-1.60 \pm 0.06) \%\ $ at $\amin = (9.1 \pm 0.1)^{\circ}$, and $\ainv = (20.2 \pm 0.1)^{\circ}$ with slope $h = (0.26 \pm 0.03) $ percent per degree. 

When comparing data, our measurements are generally in good agreement with some previous measurements, but diverge significantly with others (sometimes depending on aperture size). Overall, there is some scattering of values around the calculated best fit, which may be attributed to a number of reasons: (i) a varying state of cometary activity, especially for measurements taken at different heliocentric distances, and hence, a difference in the dust particle properties that dominate the coma, (ii) the natural evolution and aging of the comet in different apparitions, (iii) the use of different aperture sizes when measuring the polarisation, and (iv) the use of different instrumentation, wavelength domains and data reduction techniques by various groups.

First looking at the observations corresponding to the 2015-16 apparition, our measurements are in good agreement with those by \cite{kwon2022} (measured in an aperture corresponding to 1 000 km, $r \sim 2.3$ AU)---this is expected, since the observations of this group were performed at a very similar time to our 2016 observations. The measurements by \cite{rosenbush2017} are also in good agreement with our calculated best fit, but only when comparing their measurements taken in aperture sizes corresponding to a diameter of 10 000 km (which are those presented in the plot, $r \sim 1.8$ AU) or larger. Their measurements corresponding to smaller aperture sizes lead to difference of up to $\sim 4$ percentage point (p.p.)---this is due to the sharp variations in polarisation found in their polarisation maps, which will be discussed at a later point in this work. The post-perihelion measurements by \cite{hadamcik2016} taken in an aperture diameter of 10 000 km ($r \sim 1.8$ AU) are also in good agreement with our fit, while the pre-perihelion measurement with aperture diameter 1200 km ($r \sim 3$ AU) differs significantly from our fit. This difference may be attributed to the smaller aperture size chosen or to the different perihelion distance. Looking at the 2009-10 apparition, measurements by \cite{stinson2016} (aperture diameter of 10 000 km, $r \sim 3.5$ AU) are in good agreement with our own. On the other hand, the \cite{hadamcik2010} measurements (in aperture diameter 6000 km, $r \sim 1.4$ AU) differ significantly from our measurements. This difference cannot be attributed to the choice of aperture size, since their measured polarisation does not vary significantly with aperture size in most cases. It is possible this difference is due to the pre- vs post-perihelion state of the comet. Finally, looking at the 1982-93 apparition, measurements by \cite{chernova1993} (within aperture with 60 000 km diameter, $r \sim 2.4$ AU) and \cite{myers1984} (with approximate aperture diameter of 10 000 km, $r \sim 1.4$ AU) are in relatively good agreement with our own, except for those by \cite{myers1984} at small phase angles, which present a deeper NPB for 67P than more recent measurements. Some of these differences may be attributed to differences in the observations/data reduction method, while others may be attributed to intrinsic differences due to the natural evolution of the comet, including changes according to whether the comet is in the pre- or post-perihelion stage of its orbit. 

\subsubsection{Pre- vs Post-Perihelion}
Previous studies of 67P have shown evidence of a change in polarimetric behaviour before and after perihelion. \cite{hadamcik2016} found a significant difference in the fits of the pre- and post-perihelion curves at phase angles above 30\textdegree and only a slight difference at smaller phase angles. \cite{kwon2022} found a marginal difference between the fits of the post-perihelion and that of all data together. Both authors attributed these differences in polarimetric behaviour to the different characteristics of dust pre- and post-perihelion. We investigate this finding with the addition of our new observations, splitting all data into either pre- (shown in grey) or post-perihelion (red) groups, and measured the fit of each. This is shown in the right panel of Fig. \ref{fig:phasecurve}. 

We find a small difference between the polarimetric behaviour of the pre- and post-perihelion data sets. According to the fits and their errors, the pre-perihelion data set is characterised by a much deeper NPB, with minimum of $\pmin =  (-2.25 \pm  0.33) \%$ at $\amin  = (9.1 \pm  0.1) ^{\circ}$ and $\ainv  = (21.4 \pm  0.1)^{\circ}$ with slope $h = (0.31 \pm 0.02$) percent per degree, compared to the post-perihelion data set which is characterised by $\pmin = (-1.47 \pm  0.08) \%$ at $\amin  = (9.2 \pm  0.1)^{\circ}$ and $\ainv = (20.2 \pm  0.1)^{\circ}$ with $h = (0.244 \pm 0.328$) percent per degree. Although this finding provides marginal evidence of the evolution in dust properties throughout the different stages of the comet's orbit, additional pre-perihelion observations at small phase angles are needed to confirm this conclusion. It must be noted that the uncertainties for the fit values above reflect only the formal $1\sigma$ uncertainty of the fit. The true uncertainties should be much larger, especially due to the lack of pre-perihelion data points in the NPB. 

%Again, the clear difference found between the two curves supports the implication of dust property evolution throughout the perihelion passage. However, the lack of pre-perihelion data points in the NPB prevents any further in-depth analysis or solid conclusions to be achieved, and the need for more observations in this range remains. 
\begin{figure}
\begin{center}
    \centering
    \includegraphics[width=7cm,trim={0.0cm 0.0cm 0.0cm 0.0cm},clip]{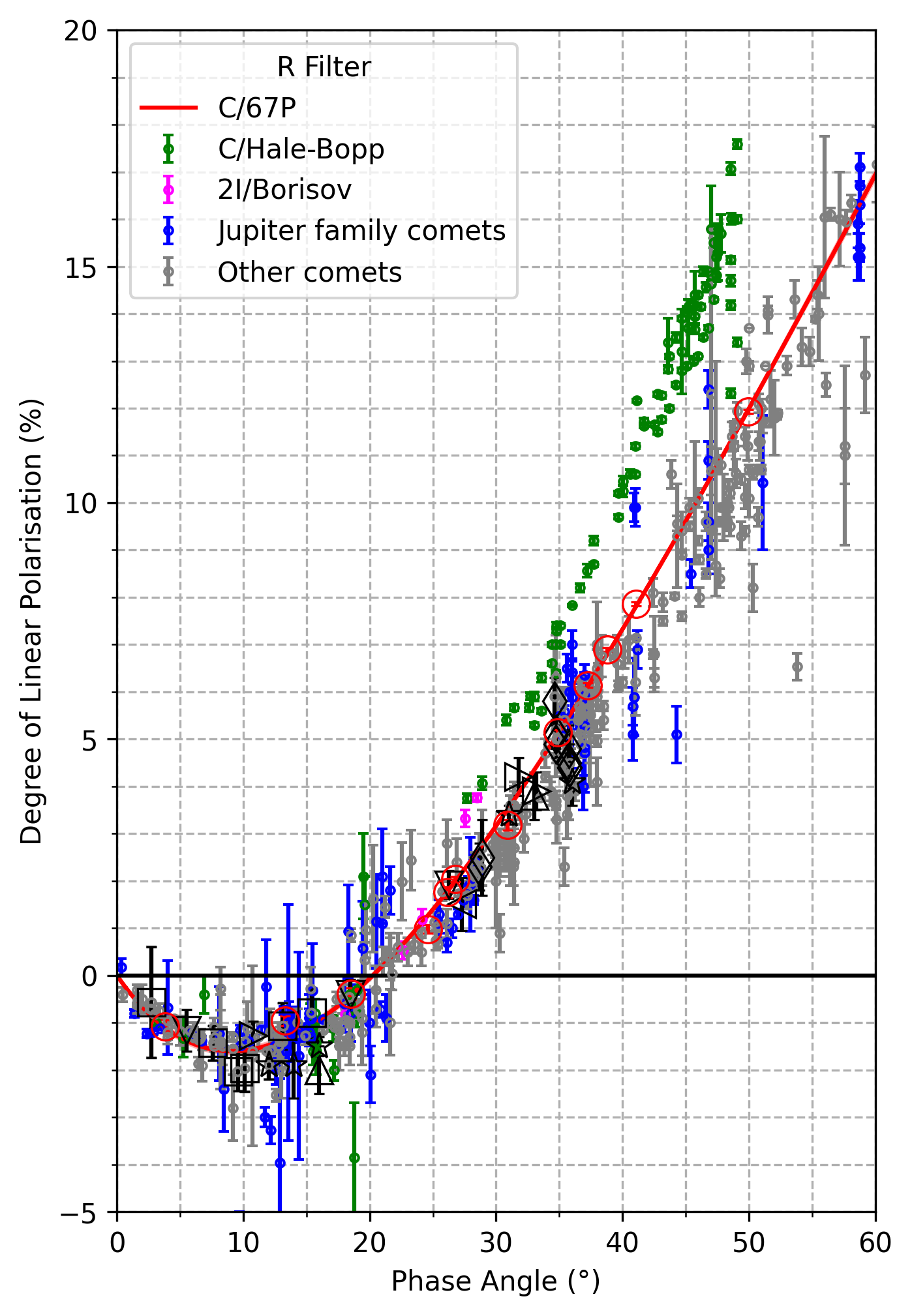}
    \caption{The polarimetric phase curve of 67P compared to those of various other comets. The red data points are the new 67P measurements presented in this work, and the red curve is the best fit of all polarimetric data of 67P (as shown in the centre panel of Fig \ref{fig:phasecurve}).}
    \label{fig:compare67P}
\end{center}
\end{figure}

\subsubsection{Comparison with Other Comets}
Finally, we compared the polarimetric phase angle dependence of our data of 67P to that of other comets recorded in literature---see Fig. \ref{fig:compare67P}. For this, we have compiled literature data from the NASA/DBCP archive \citep{kiselev2017} obtained with filter domains similar to that of the R filter. We have avoided using data taken with filters designed to cover gas emission bands for our comparison. Despite this, the presence of molecular bands may still affect measurements with broadband filters. To minimise these effects, we chose the data point with the smallest aperture in the case where the polarisation of the comet was measured in multiple aperture sizes. This ensures the measured value is most representative of the inner most coma where the highest values of polarisation are usually observed. Further, we have excluded data points with very large error bars. We have distinguished the data of polarimetrically peculiar comets C/1995 O1 (Hale-Bopp) \citep{tanga1997,hadamcik2003b,hadamcik2003c} and 2I/Borisov \citep{bagnulo2021} to that of other Jupiter family comets (JFCs) and various other comets. The red curve is the best fit 67P (all data, as presented in the centre panel of Fig. \ref{fig:phasecurve}). Our new observations confirm that the polarimetric behaviour of 67P is similar to that of the majority of other JFC comets, and fits into to the high-polarisation comet category. The general scattering of the polarimetric data points may be attributed to a varying state of cometary activity among the comets, as well as a consequence of the use of different instrumentation and wavelength domains by various groups.

 \subsection{Polarimetric spectral gradient} \label{sect_spectral}
 \begin{figure}
 \centering
 \includegraphics[width=6.7cm,trim={0.0cm 0.0cm 0.0cm 0.0cm},clip]{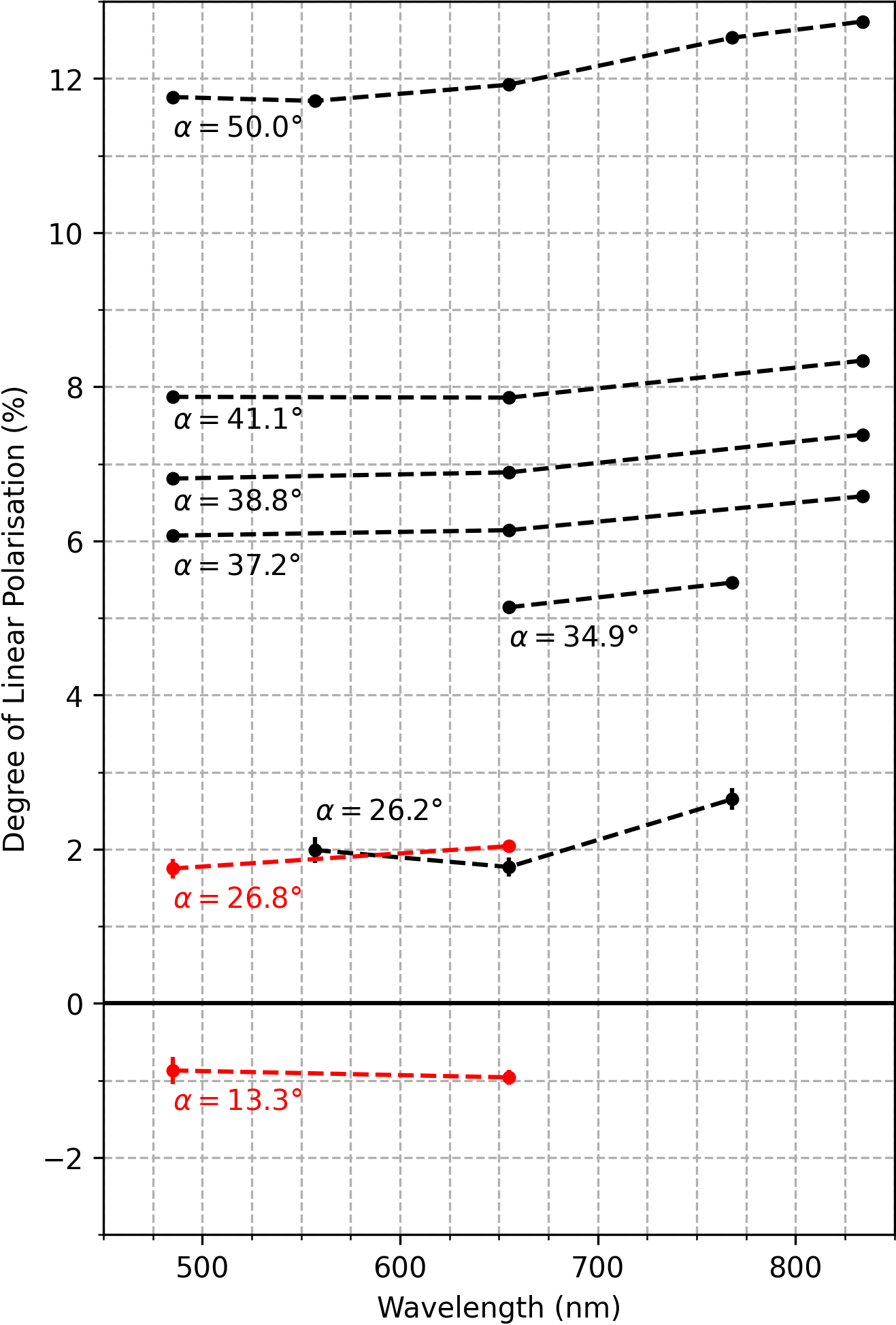}
 \caption{\label{fig:polwaves} Broadband polarimetry of 67P obtained at various phase angles as shown in the plot. The red symbols and lines refer to the post-perihelion data obtained in 2016, while the rest are the pre-perihelion measurements taken in 2021.}
 \end{figure}

By using various filters in our measurements, we can examine the variation in polarisation with wavelength at various phase angles. In Fig. \ref{fig:polwaves}, we plot our FORS2 measurements from both 2016 and 2021 as a function of wavelength. This general behaviour is comparable to previous measurements of various other comets --- e.g., see \cite{chernova1993,kiselev2015}.

It is also of interest to measure the polarimetric spectral gradient (PSG). Following, e.g. \cite{bagnulo2021}, we define:
 \begin{equation}
 \left.\frac{{\rm d}\,\vert \pq (\lambda)\vert}{{\rm d}\,\lambda} \right \vert_{\lambda=\lambda_0} \simeq {\rm PSG}(\lambda_1,\lambda_2) = \frac{\vert\pq (\lambda_2)\vert - \vert\pq (\lambda_1)\vert}{\lambda_2-\lambda_1} \\
 \label{eq:polspecgrad}
 \end{equation}
 where
 \begin{equation}
 \lambda_0 = \frac{\lambda_1 + \lambda_2}{2} \; ,
 \end{equation}
 where $\vert \pq \vert$ is the absolute value of \pq. According to this equation, PSG is positive when the degree of polarisation (absolute value) increases with wavelength, regardless of its direction. Around zero polarisation, however, the PSG is not defined. In Table \ref{tab:polcol}, we record the polarimetric spectral gradient of 67P for various pairs of filters. We cannot further refine our PSG measurements since our measurements were obtained mostly in broadband filters. 
 
\begin{figure}
 \centering
 \includegraphics[width=8cm,trim={0.0cm 0.0cm 0.0cm 0.0cm},clip]{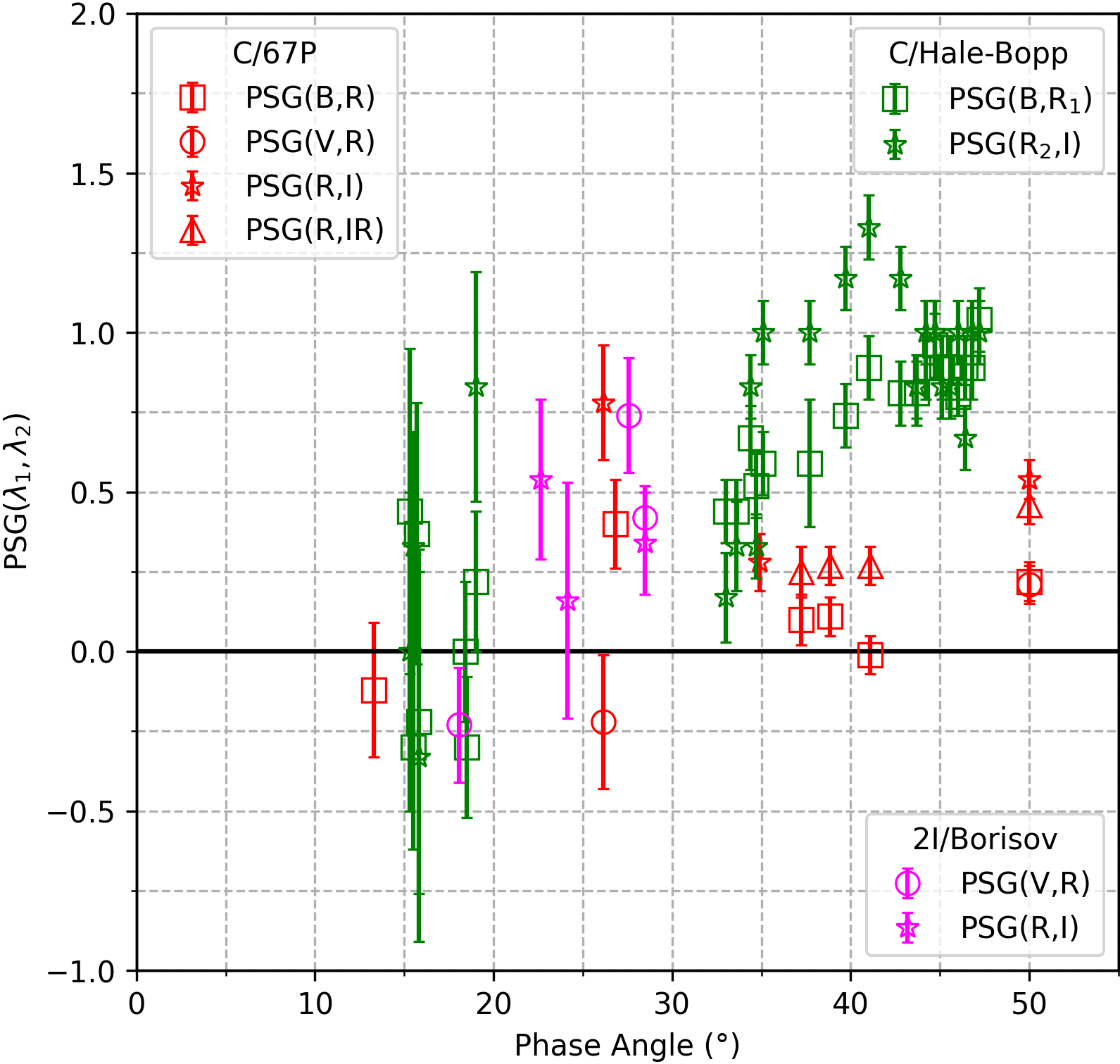}
 \caption{\label{fig:polcol} Polarimetric spectral gradients (PSG) of 67P as a function of phase angle, compared to those of Hale-Bopp and 2I/Borisov. For 67P, (B, V, R, I, IR) refer to measurements taken with filters centred around wavelengths (485, 557, 655, 768, 834) nm, respectively. For Hale-Bopp, (B, $\rm{R_1, R_2}$,I) are filters centred around (485, 620, 670, 730) nm, and (V, R, I) are (557, 655, 730) nm for 2I/Borisov. We note that the polarisation is directed along the scattering plane at phase angles values $\le 21^\circ$, while it is directed along the perpendicular to the scattering plane at phase angles $ \ge 20^\circ$.}
 \end{figure}

In Fig. \ref{fig:polcol}, we plot our PSG measurements of 67P as a function of phase angle, and compare them to those of comets C/1995 O1 (Hale-Bopp) and 2I/Borisov. The significance of our comparison is limited by the fact that polarimetric measurements are typically obtained at different phase angles, as well as the fact that different filters are used. In this case, we have estimated the PSG values of Hale-Bopp using measurements taken with filters centred at $\lambda_1 = 485$ nm and $\lambda_2 = 620$ nm for PSG(B, $\rm{R_1}$), and 670 and 730 nm for PSG($\rm{R_2}$, I). For 2I/Borisov, we have used measurements taken with filters centred at 557 and 655 nm for PSG(V, R), and 655 and 730 nm for PSG(R, I). Comet 67P shows a mildly positive polarimetric spectral gradient, typical of most comets, but somewhat flatter compared to Hale-Bopp and 2I/Borisov. 

\begin{table}
\begin{center}
\caption{\label{tab:polcol} Polarimetric spectral gradient (PSG) of 67P according to Eq. \ref{eq:polspecgrad}. Here, (B, V, R, I, IR) refer to measurements taken with filters centred around wavelengths (485, 557, 655, 768, 834) nm, respectively.}
\begin{small}
\begin{tabular}{cr@{$\pm$}lr@{$\pm$}lr@{$\pm$}lr@{$\pm$}l}
    \hline\hline
     & \multicolumn{8}{c}{PSG($\lambda_1,\lambda_2$)} \\
     $\alpha$ &  \multicolumn{2}{c}{(B,R)} & \multicolumn{2}{c}{(V,R)} & \multicolumn{2}{c}{(R,I)} & \multicolumn{2}{c}{(R,IR)} \\
      $(^{\circ})$ & \multicolumn{2}{c}{$\% /100$ nm} & \multicolumn{2}{c}{$\% /100$ nm} & \multicolumn{2}{c}{$\% /100$ nm} & \multicolumn{2}{c}{$\% /100$ nm}  \\
    \hline
26.8 & 0.4 & 0.14 &  \multicolumn{2}{c}{-} &  \multicolumn{2}{c}{-} &  \multicolumn{2}{c}{-} \\
13.3 & -0.12 & 0.21 &  \multicolumn{2}{c}{-} &  \multicolumn{2}{c}{-} &  \multicolumn{2}{c}{-} \\
\hline
26.2 &  \multicolumn{2}{c}{-}  & -0.22 & 0.21 & 0.78 & 0.18 &   \multicolumn{2}{c}{-} \\
34.9 &  \multicolumn{2}{c}{-} &  \multicolumn{2}{c}{-}& 0.28 & 0.09 &  \multicolumn{2}{c}{-} \\
37.2 & 0.1 & 0.08 &  \multicolumn{2}{c}{-} &  \multicolumn{2}{c}{-} & 0.25 & 0.08 \\
38.8 & 0.11 & 0.06 &  \multicolumn{2}{c}{-} &  \multicolumn{2}{c}{-} & 0.27 & 0.06 \\
41.1 & -0.01 & 0.06 &  \multicolumn{2}{c}{-} &  \multicolumn{2}{c}{-} & 0.27 & 0.06 \\
50.0 & 0.22 & 0.06 & 0.21 & 0.06 & 0.54 & 0.06 & 0.46 & 0.06 \\
    \hline
\end{tabular}
\end{small}
\end{center}
\end{table}

%%%%%%%%%%%%%%%%%%%%%%%%%%%%%%%%%%%%%%%%%%%%%%%%%%

\subsection{Imaging Maps}
\begin{figure*}
\begin{center}
    \centering
    \includegraphics[width=15cm,trim={0.0cm 0.0cm 0.0cm 0.0cm},clip]{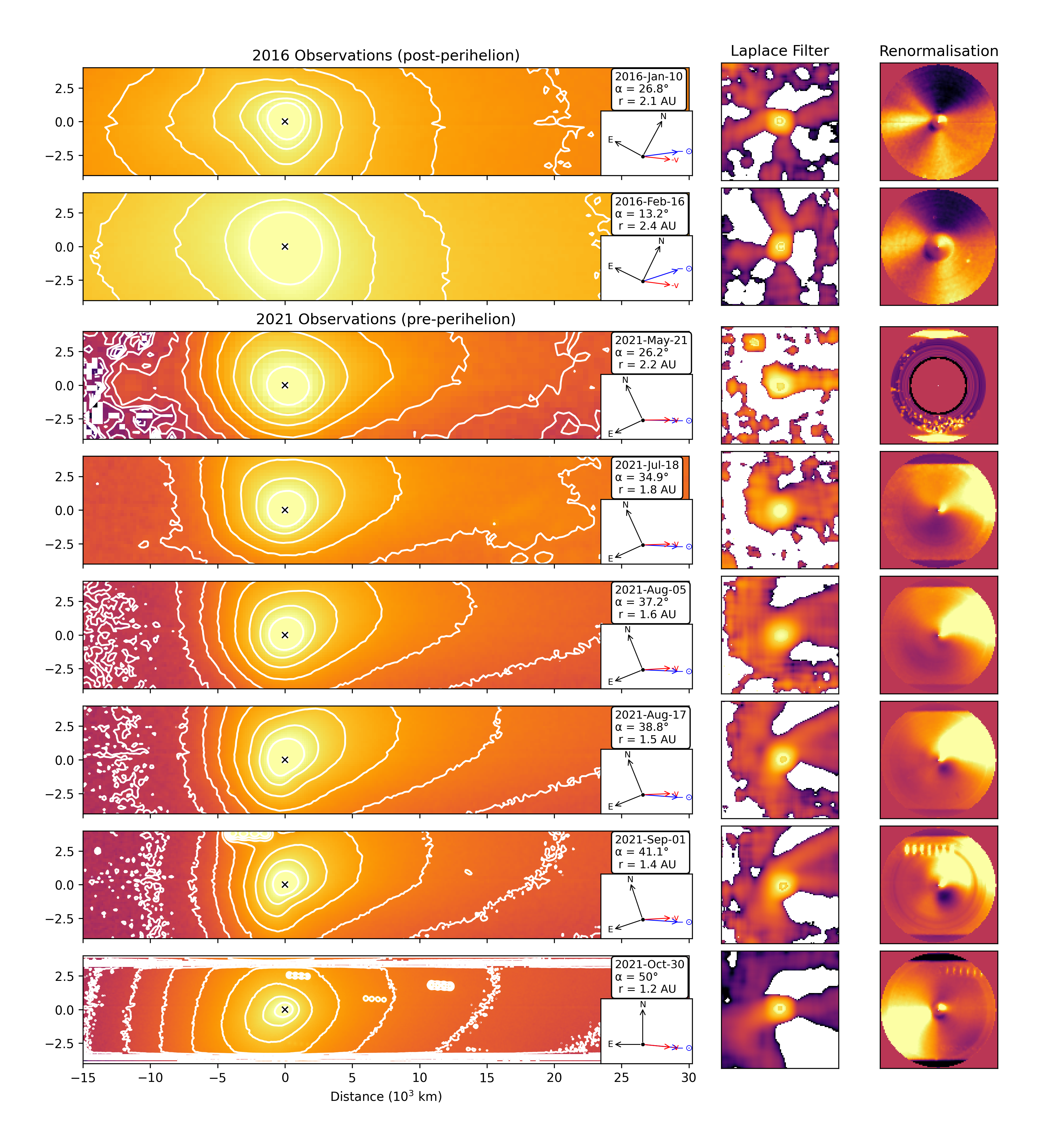}
    \caption{Original and enhanced imaging maps of 67P in R-filter taken with FORS2. Each row represents one observing epoch. The date, phase angle ($\alpha$), and heliocentric distance (r) of each epoch is given in the text box in the left panels.  The north (N), east (E), anti-solar direction ($\odot$) and anti-velocity vector (-v) of the comet from the point of view of the observer are also shown in the left panels. \textit{Left:} imaging maps, i.e. deep images, displayed with logarithmic scale normalised to one with isophotes. \textit{Middle:} imaging maps with applied Laplace filter, arbitrary scale. \textit{Right:} imaging maps with applied renormalisation filter, arbitrary scale.}
    \label{fig:IntMaps}
\end{center}
\end{figure*}

In Fig. \ref{fig:IntMaps}, we have plotted the imaging maps, as well as enhanced images, of our VLT data in the R-filter. Each row in this plot corresponds to an observing epoch, details of which are given in the text boxes of the left panels. The north (N), east (E), anti-sunward ($-\odot$), and anti-velocity ($-v$) directions are given in the left panels. 

The left panels display the deep-imaging maps, plotted in a logarithmic scale and scaled to the same distance scale for viewing purposes: up to 30 000 km along the tail and 4 000 km in the perpendicular direction. The photocentre of the comet has been marked with an X in these maps and we have also plotted isophotes to highlight any variations in intensity, which may be indicative of regions of cometary activity. The two right panels display the acquisition images of each epoch enhanced with the Laplace filter and radial renormalisation method. Coma structures found on the anti-solar hemisphere (mostly western and north-western directions) are related to the tail of the comet, while those in the sunward hemisphere (mostly eastern and south-eastern directions) may indicate enhanced activity on the nucleus of 67P. Given the wavelength bandpass of the R filter used, it can be assumed that the identified structures are caused by sunlight reflected by dust. This is true except for features caused by the passing of background stars, which appear as bright, sometimes elongated, spots as found in epochs 2021-Sep-01, and 2021-Oct-30.

Boehnhardt et al. (2024, in preparation) presents a comprehensive analysis of the dust activity, coma and tail structures, and stability of the rotation axis based on our FORS2 observations, as well as observations taken at the Observatorio de Sierra Nevada in Spain, operated by the Instituto de Astrofísica de Andalucía (IAA). In this context, we present a brief analysis to compare to our polarimetric measurements. 

Comet 67P displayed an extended coma and tail over the entire observing epoch. First, we will look at the 2016 maps. Circular or oval isophote patterns may be considered as an indicator of somewhat uniform cometary activity across the surface nucleus. Instead, ``bumps" in the isophote patterns highlight areas of enhanced dust reflectivity, indicative of localised coma features. In particular, the isophotes in the 2016 maps (upper two panels) show protrusions in the eastern hemisphere and south-western quadrant. The coma enhancement techniques applied to the acquisition images of these epochs, shown in the panels to the right, reveal two or three coma patterns resembling fan-like structures. These enhanced images also make it possible to measure the position angle of these structures. From the north direction anti-clockwise, the position angle (PA) of the jet-like structures in 2016-Jan-10 are $\sim 76,\ 118$ and $208^{\circ}$, while the tail is at PA $\sim 303^{\circ}$. In 2016-Feb-16, the structures are at PA $\sim 115$ and $216^{\circ}$ and the tail at $302^{\circ}$. Other than changing position due to the viewing geometry of the comet, these features do not vary significantly between observations. These measurements are comparable to those by \cite{hadamcik2016}, \cite{boehnhardt2016} and \cite{rosenbush2017}. In particular, the overall geometry and orientation of the fans resemble the X2 pattern described in the latter study, which is characterised by three fan-like structures in addition to the tail.

Next, looking at the maps of the 2021 epochs, a global increase in activity is evident as the heliocentric distance of the comet (given as $r$) decreases from 2.2 to 1.2 AU. On the other hand, the relative Earth-comet distance ($\Delta$) decreases from 2.3 AU to 0.4 AU throughout the observing epoch, giving us better spatial resolution of the coma, as well as a higher S/N ratio due to the increase in brightness. The evolution of certain features can be followed as the comet approaches perihelion. The northern-most jet-like feature, which first appears in 2021-May-21 around PA $40^{\circ}$ (from northern direction, anti-clockwise), gradually fades and migrates to PA $30^{\circ}$ throughout the observing epoch. Another jet-like structure appears on 2021-Aug-05 at PA $125^{\circ}$, which becomes more prominent and moves to PA $105^{\circ}$ as the comet approaches perihelion. The tail remains around PA $\sim 250-270^{\circ}$ throughout the observations. The jet-like structure and tail geometry of 67P in the 2021 epoch, in particular in the observations closest to perihelion, resemble the X1 pattern, characterised by two main jet structures and a dust tail, described by \cite{boehnhardt2016} of the immediate post-perihelion phase of 67P in 2015. Overall, our observations show that the localised enhanced activity was stronger in the northern region of the nucleus in the early 2021 observations, while activity in the southern region was absent. Later, activity in the northern region gradually faded and became dominant in the southern hemisphere as the comet approached perihelion. A similar evolution of coma structures in comet 67P is known from the analysis of earlier apparition described in \cite{vincent2013} and \cite{boehnhardt2016}. The similarity of the coma structure patterns around perihelion in 2015 and 2021 may suggest that the same regions were active during these subsequent returns of the comet towards the Sun. Whether this also implies a similar orientation of the rotation axis of the nucleus, despite the close encounter of 67P with Jupiter in late 2018, requires a deeper modelling analysis involving also coma images of the comet from the post-perihelion period in 2021.

%%%%%%%%%%%%%%%%%%%%%%%%%%%%%%%%%%%%%%%%%%%%%%%%%%

\subsection{Polarimetric Maps}

\begin{figure*}
\begin{center}
    \centering
    \includegraphics[width= \textwidth]{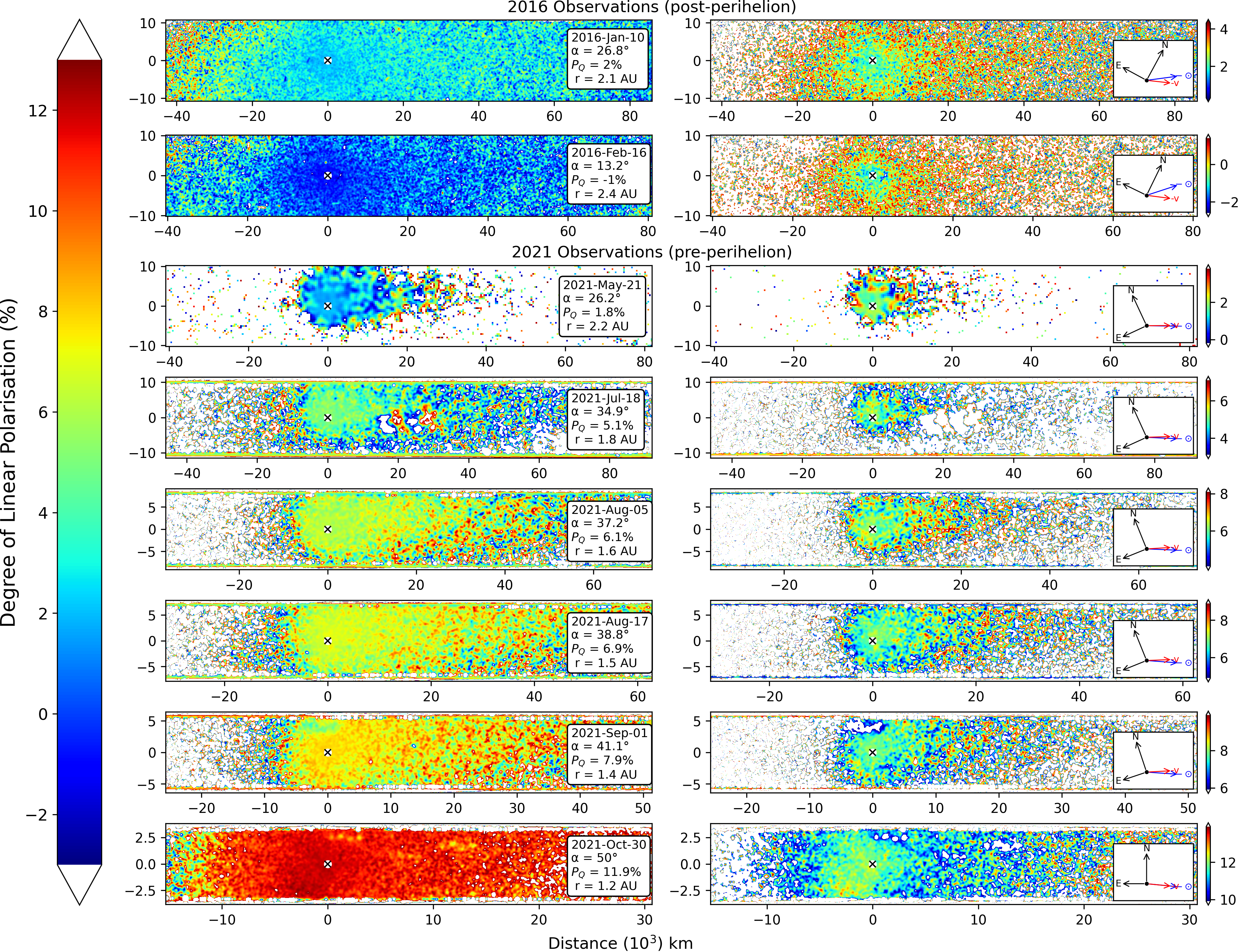}
    \caption{\pq\ polarimetric maps of 67P in R-filter taken with FORS2, with a Gaussian convolution filter applied. As in Fig. \ref{fig:IntMaps}, each row represents one epoch, the details of which are given in the text boxes. \textit{Left:} polarimetric maps with scale $(-3,13)\%$. The colour of each pixel represent the value of the degree of linear polarisation, as indicated in the colourbar to the left. For better visualisation, the colour of all pixels whose value falls outside the range  $(-3,13)\%$ has been set to white and can be considered noise. \textit{Right:} polarimetric maps with dynamical scale --- $\pq \pm 2\%$, where \pq\ is the value measured from aperture polarimetry indicated in the text box. Again, pixels whose value falls outside this range has been set to white.}
    \label{fig:PolQMap}
\end{center}
\end{figure*}

In Fig. \ref{fig:PolQMap}, we have plotted the \pq\ polarimetric maps of our VLT data in the R-filter. Again, each row in this plot corresponds to an observing epoch and the directional vectors are given in panels to the right. 

In the left panels, the colour of each pixel represents the value of polarisation as given in the colourbar to the left. The value measured from aperture polarimetry has been given as \pq\ in the text box. Taking this value into account, the polarimetric maps have been plotted again in the right panels, but with a scale of $\pq \pm 2\%$. For example, $\pq = 2\%$ for the case of 2016-Jan-10 (first row), and therefore, the scale has been set to the range $0$ to $4\%$ for this particular map. Any pixel with a value outside this range is given in white. The reason for plotting the maps in this manner is to enhance the contrast and emphasise any small scale variability of polarisation. 

The polarimetric maps show that the cometary coma and the observed region of the tail appear to have an almost homogeneous distribution of polarisation around the average \pq\ value measured from aperture polarimetry in each epoch. No strong features or gradients in polarisation are apparent, despite the evident jets detected in the imaging maps. This uniformity is reflected in the aperture polarimetric measurements as shown in Fig. \ref{fig:growthcurves}. Throughout the 2021 observing epochs, the overall area of polarised light of the coma and tail becomes larger and more dense. This can be attributed to the increase in cometary activity as 67P approaches perihelion and as more of the ejected particles are pushed in the anti-solar direction due to solar radiation pressure. Similarly, the area of polarised light is much larger in the 2016 observing epoch, likely due to the peak of cometary activity being reached after perihelion (2015-Aug-13). Some false ``features" are present in a number of these maps: a ``gap" appears in the tail of the map of 2021-Jul-18, and some ``spots" appear in the upper coma of epochs 2021-Sept-01 and 2021-Oct-30, respectively. These are due to the passing of background stars. Since the maps are made by combining multiple frames taken seconds or minutes apart, these stars appear as "moving spots" in the maps. They should therefore be ignored in the interpretation of the maps.

Measurements and models of 67P dust coma seem to indicate that the optical cross section is dominated by large particles, in excess of $\sim$10 $\mu$m. The analysis of all the available 67P dust coma, tail, and trail images of several comet apparitions until 2010 were analysed by \cite{fulle2010} in their 67P dust environment model for the Grain Impact And Dust Accumulator (GIADA) on board Rosetta. The results indicated that the dust cross section was dominated by mm to cm-sized particles. During the Rosetta orbit (2015 perihelion), \cite{moreno2017} analysed a large data set of ground-based dust tail images, from 4.5 AU pre-perihelion to 3 AU post-perihelion, finding that a minimum particle size of 20 $\mu$m was compatible with all the observations. On the other hand, the dust scattering phase function of 67P, observed by the OSIRIS instrument on Rosetta \citep{bertini2017}, which is characterised by a strong backscattering enhancement, has also been interpreted by large absorbing and porous grains, in the radii range 5-100 $\mu$m, by \cite{markkanen2018}. In addition, Rosetta/VIRTIS-H thermal spectra modelling needed relatively compact particles with minimum sizes of 10 $\mu$m (combined with 25\% fractals in number) \citep{bockelee-morvan2017}. The atomic force microscope MIDAS on Rosetta has detected mostly large ($>$10 $\mu$m) agglomerated particles with moderate packing or porous, with very scarce individual small particles. In fact, no individual particles smaller than 1 $\mu$m have been detected \cite{mannel2019}. The instrument COSIMA on Rosetta imaged a large number of particles ranging from 10 $\mu$m up to 1 mm, suggesting that large particles dominate the scattering features of the coma, very likely being the same agglomerates observed by MIDAS \citep{hilchenbach2016}. In line with those determinations, direct measurements using  the Micro Balance Subsystem of GIADA instrument on board Rosetta confirm that submicrometre- to micrometre-sized  particles do not provide a substantial optical scattering in the coma of 67P with respect to the scattering caused by millimetre-sized particles \citep{della-corte2019}.

All of those lines of evidence point toward a dust coma dominated by grains in the geometric optics regime, for incident wavelengths in the visible regime. Since all the polarisation phase curves in such a regime converge to a single curve irrespective of the particle size, for both compact and porous absorbing particles at least to $60^{\circ}$ of phase angle \citep[see][]{markkanen2018}, that would explain quite naturally the absence of structures or gradients in the polarisation maps. Those gradients would otherwise show up if the size distribution peaked in the submicrometre- or micrometre-sized particle region as a consequence of radiation pressure pushing those smallest particles towards the outermost regions of the tails.  

The homogeneous spread of polarisation found in our maps is not in agreement with some previous observations. In particular, the polarimetric maps of 67P by \cite{rosenbush2017} created from observations taken in November and December of 2015, as well as April of 2016 show strong variations in polarisation with distance from the comet centre, in the order of an $\sim 8$ p.p. change in degree of linear polarisation. Similarly, their aperture polarimetric measurements show strong variations with aperture size. In contrast, both our 2016 and 2021 polarimetric maps, the former of which should be closely comparable to the study mentioned, show a uniform spread of polarisation with little variation with distance from the comet. 

The discrepancies between our polarimetric maps and those presented in \cite{rosenbush2017} could be explained with a number of reasons. On the one hand, there are significant differences in the dimensions of the polarimetric maps: our maps image only the central region of the coma while \cite{rosenbush2017} obtain a much wider FoV of the comet. However, even in instances where the central regions of our polarimetric maps coincide with those of this article, we have not observed similar gradients in polarisation value. On the other hand, considering the 2016 apparition only, the comet was observed under different observing conditions: \cite{rosenbush2017} observed the comet at heliocentric distances r=1.61, 1.84, and 1.81 AU with phase angles $\alpha = 33.2^{\circ}, 31.8^{\circ}$, and $10.4^{\circ}$, respectively, while we observed at r=2.09 and 2.37 AU with $\alpha = 26.8^{\circ}$ and $13.3^{\circ}$. However, this does not explain the fact that we detect features in our intensity maps but not in our polarimetric maps. Further, it is possible that the activity levels of the comet varied between epochs, leading to differing polarisation maps. A final possibility could be differences in the overall observational and/or data reduction methods. Despite considering these potential factors, we are still uncertain of the exact reason for the differences in polarimetric maps.

Although our findings deviate from those of previous studies, we remain confident in our results under the following simple consideration. The value of each pixel in a polarimetric map is calculated by the difference of the fluxes of the individual pixels of the original frames. It is extremely unlikely that a fault in the data reduction, e.g. the misalignment of polarimetric frames, can cancel an existing signal and produce a homogeneous map. In other words, it is unlikely that the differences measured from adjacent pixels are similar among themselves if the source itself is not polarimetrically homogeneous. With the exception of very trivial errors, a fault in the data reduction method will more likely result in the variable polarisation signal rather than a smooth map. We will further explore this matter in the following section.

\section{Qualitative comparison with previous polarimetric studies} \label{sec_misalignment}

\begin{figure}
    \centering
     \subfigure[]{
        \includegraphics[width=0.8\linewidth]{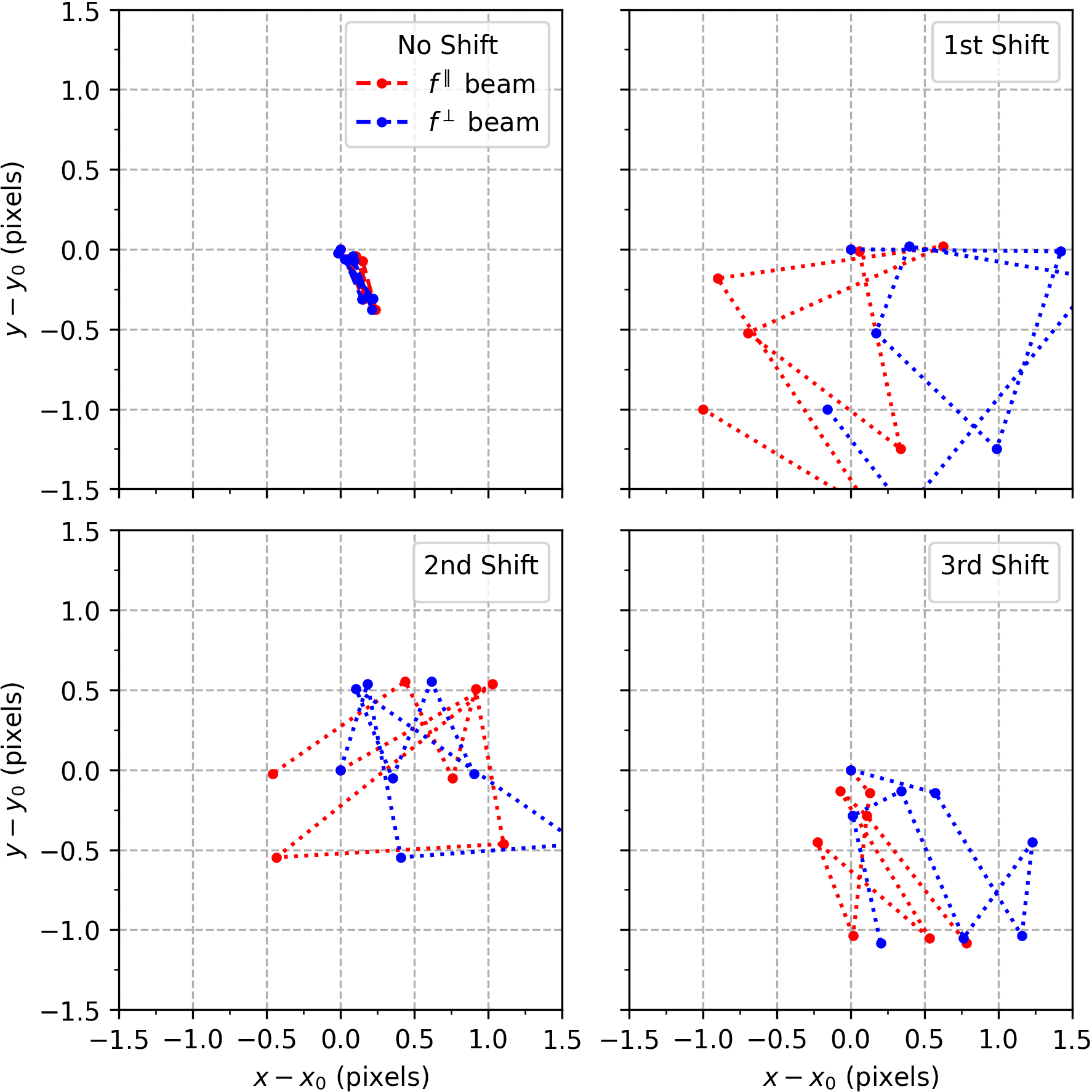}
        \label{fig:misalign1}}
     \subfigure[]{%{0.3\textwidth}
        \includegraphics[width=0.8\linewidth]{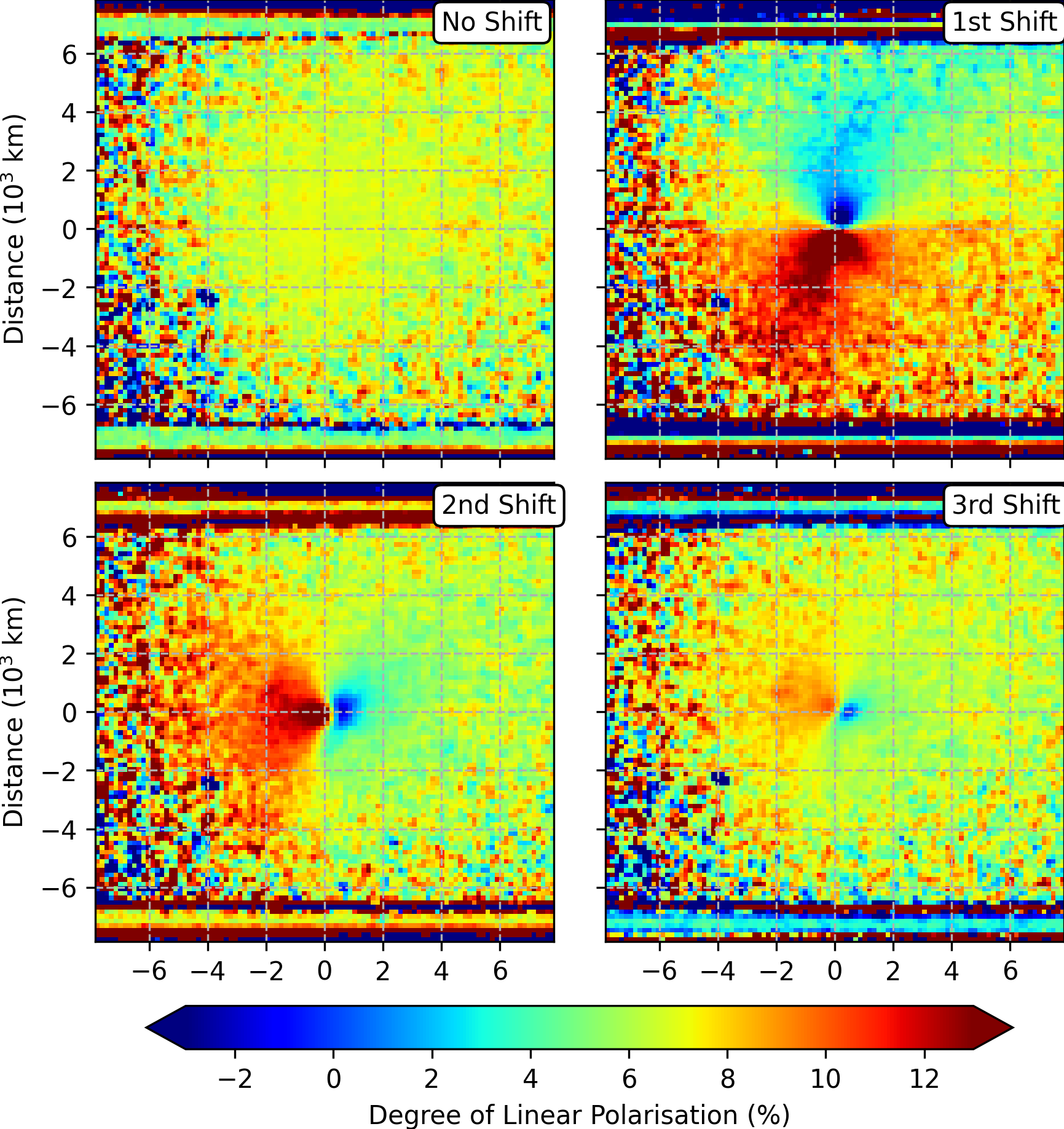}
        \label{fig:misalign2}}
         \subfigure[]{%
        \includegraphics[width=0.8\linewidth]{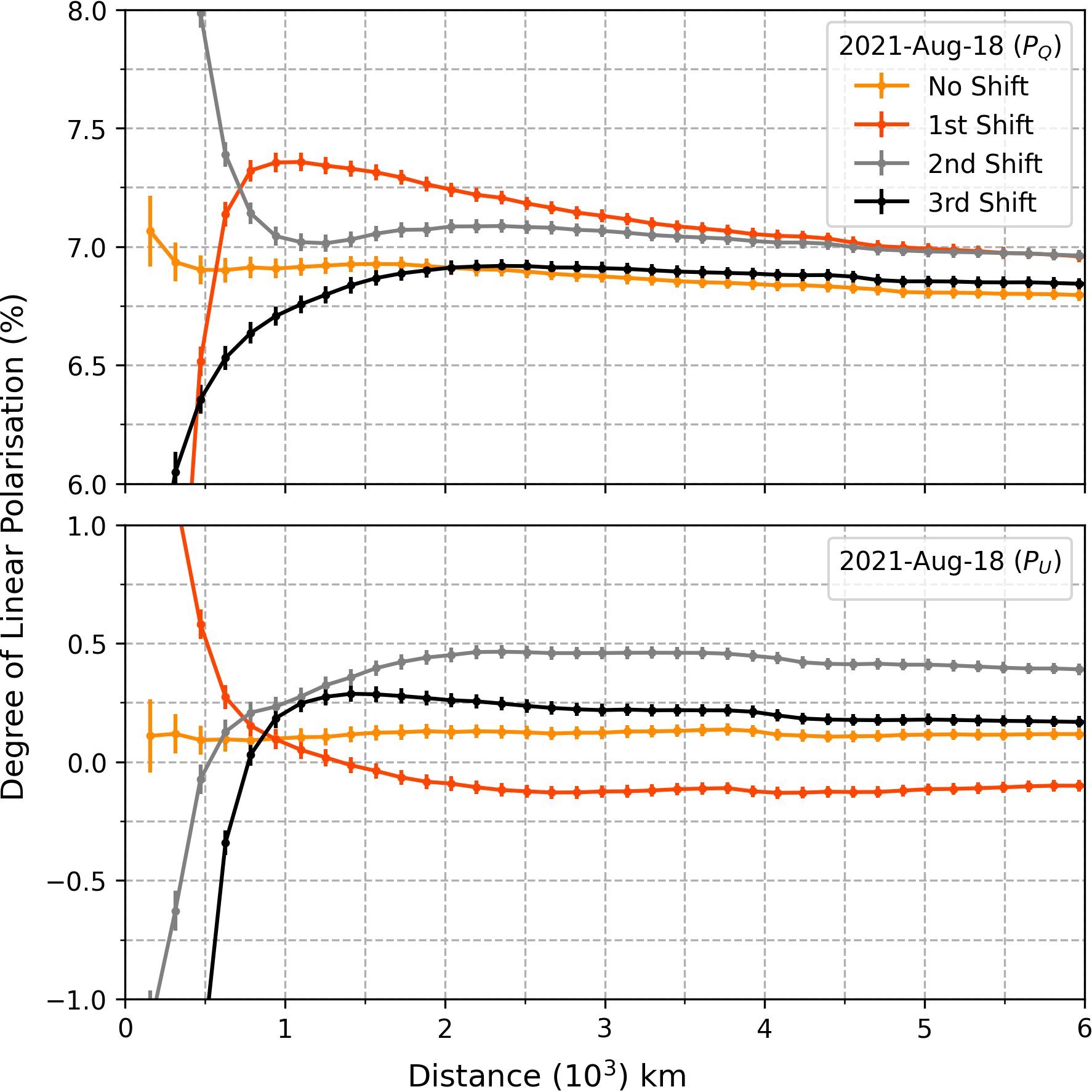}
        \label{fig:misalign3}}
    \caption{Misalignment experiment as described in the text, performed using R-filter FORS2 data of epoch 2021-Aug-17. The ``No Shift" data/panels refer to the original data without introducing deliberate misalignments.}\label{fig:misalign}
\end{figure}

In literature on the topic, polarimetric maps of comets have been used to reveal regions of varying properties of particles in the coma. Dust jets and fans have typically been reported to exhibit an increased level of polarisation compared to the surrounding coma, while rapid variations in polarisation within the coma have been reported as a sign of fragmentation and/or sublimation of the ice layers within the individual dust grains \citep{hadamcik2003c}. This is thought to be due to the concentration of smaller particles due to ejection and/or the alignment of the ejected particles affected by gas drag. Similarly, polarimetric maps show that particles ejected from nucleus fragmentation \citep[e.g., C/1999 S4 (LINEAR),][]{hadamcik2003} or impact \citep[e.g., 9P/Tempel 1 after the Deep Impact event,][]{furusho2007,hadamcik2007,harrington2007} are not that different to those ejected after an outburst. 

Often, comparisons between intensity and polarimetric maps show that regions of higher polarisation correlate well with areas of enhanced dust emission: close to the nucleus and in the jets. However, upon close visual inspection, this is not always the case. For instance, comparing the intensity and polarimetric maps of C/1995 O1 (Hale-Bopp) \citep{tanga1997,hadamcik2003b,hadamcik2003c}, it appears that areas of prominent activity do not always correlate well with features found in the polarimetric maps. Instead, a region of high-contrast polarisation is typically found near the photocentre. We have observed a similar inconsistency in the polarimetric maps of comets C/1990 K1 (Levy) and 47P/Ashbrook-Jackson \citep{renard1992,renard1996,hadamcik2003c}. According to the authors, the features in the polarimetric maps were not permanent, and their evolution in position and polarisation varied from observation to observations, despite few changes measured in the intensity maps. In other words, the correlation between jets identified in intensity maps and regions of high/low polarisation in polarimetric maps appears to be inconsistent.

While reports of a uniform spread of polarisation across cometary comae are infrequent in older data sets, they appear more common in some recent data. Polarimetric maps of comet C/2017 K2 by \cite{zhang2022}, generated with high spatial resolution HST data, display a featureless and almost homogeneous spread of polarisation across the coma and tail, despite a presence of a jet-like fan evident in the intensity image. Additionally, this paper revisits HST polarimetric measurements of comet ISON (C/2012 S1) conducted by \cite{hines2013}, who found an unusually sharp, positively polarised region close to the nucleus, which they associated with an extended jet-like feature. After excluding a frame affected by a minor $\sim 1-2$ pixel tracking error, the positively polarised peak disappeared. This is a prime example of a spurious polarisation measurement due to faulty data. As well as measuring a higher level of polarisation compared to most comets, \cite{bagnulo2021} found a homogeneous spread of polarisation in the coma of comet 2I/Borisov, which the author's attributed to the pristine nature of the interstellar visitor.

Another interesting case is that presented by \cite{shinnaka2023}, who found that the intensity maps of comet 21P/Giacobini-Zinner showed an elongated coma structure along the solar-antisolar direction, while the polarimetric maps showed an almost uniform spread of polarisation, with no evidence of jets or arc structures in the inner coma. This was unexpected since the polarimetric maps by \cite{chornaya2020}, created from observations taken on the same night as those by \cite{shinnaka2023} and three additional nights around that date, show a significant variation in polarisation. In particular, their polarimetric maps show a high polarisation region ($\sim$ 21 \%) in the solar direction and a low polarisation region ($\sim 8\%$) in the anti-solar direction, forming a region of high-contrast polarisation with a difference of $\sim 13$ p.p. close to the photocentre. Despite the differences in polarimetric maps, however, both groups reported the very similar values from aperture polarimetric measurements in a $\sim 5000-6000$ km aperture centred at photocentre. Notably, there are two main differences between the data reduction techniques of the two groups. 

On one hand, there is the observation technique and the method of calculating the Stokes parameters. \cite{shinnaka2023} used images taken at four different wave-plate angles ($0,22.5,45,67.5^{\circ}$) to measure the polarisation, whereas \cite{chornaya2020} took images with a dichroic polarisation filter rotated at three position angles ($0,60,120^{\circ}$). On the other hand, there is the algorithm used to calculate the position of the photocentre, and hence, align the individual images when constructing the polarimetric maps. \cite{shinnaka2023} calculated the photocentre according to the position of the Gaussian peak in amplitude of the comet, while \cite{chornaya2020} determined the position using the brightest isophotes closest to the photocentre. Although both methods yield similar values with aperture polarimetry, one method results in uniform polarimetric maps, while the other produces a region of high-contrast polarisation close to the photocentre. Interestingly, \cite{shinnaka2023} employed the same data reduction method as we did in this work, regarding both the polarimetry and alignment algorithm, and obtained a homogeneous spread of polarisation in agreement with our results. In contrast, \cite{chornaya2020} used the same method as \cite{rosenbush2017}, and both groups obtain sharp variations in polarisation around the photocentre. 

Despite the differing polarimetric maps, we note that \cite{chornaya2020} and \cite{shinnaka2023} both measured similar values from aperture polarimetry. To investigate the possibility of measuring the same value from aperture polarimetry from different polarimetric maps, we conducted an experiment using our R-filter data of 17-Aug-2022. First, we added a random value between (--1,1) to the original x- and y-positions of the comet photocentre (calculated with the DAOStarFinder Python package) of each of the \fe\ and \fo\ beams of all frames (retarder waveplate angles $0, 22.5, 45^{\circ}$, etc.), thus, deliberately introducing minor misalignments. Secondly, we plotted polarimetric maps with our misaligned frames, and lastly, we performed aperture polarimetry. The results of our experiment, which we iterated three times and compared to the original results, are shown in Fig. \ref{fig:misalign}.

Fig. \ref{fig:misalign1} tracks the pixel coordinates of comet photocentre of frames $22.5-157.5^{\circ}$ relative to the first $0^{\circ}$ frame, i.e., the comet photocentre of the $0^{\circ}$ frame is at (x,y) position (0,0). The top left panel of this plot shows the original pixel coordinates, while the remaining panels show the three iterations of our experiment. Fig. \ref{fig:misalign2} shows the polarimetric maps created with these misalignments, compared to the original polarimetric map of this epoch (top left panel). In the misaligned cases, various scales of high-contrast regions are found close to the photocentre.  These plots demonstrate how spurious features are created by misaligned frames, and thus, may be misinterpreted as cometary jet- or fan-like structures. Finally, Fig. \ref{fig:misalign3} shows the polarisation measured as a function of aperture for all four cases. While measurements taken at smaller aperture sizes may differ significantly, the values tend to converge at larger aperture sizes, differing no more than a small fraction of a percentage. This is because the two negative and positive extremities of polarisation eventually average out at larger aperture sizes.  Although this experiment presents an exaggerated example of misalignments, it proves that values measured from aperture polarimetry can remain nearly identical at certain aperture sizes for both aligned and misaligned cases. In contrast, even minor misalignments can give rise to spurious features around the photocentre in polarimetric maps, which may be interpreted as jets. 

In summary, our experiment highlights the sensitivity of polarimetric maps to misalignments, despite obtaining comparable values via aperture polarimetry. It is important to note, however, that our discussion does not imply the superiority of one method over the other, regarding polarimetry or the alignment of frames, nor does it suggest the inaccuracy of previous studies. For example, we cannot conclude that the features found in the polarimetric maps of \cite{rosenbush2017} are due to misalignments, as their shape do not resemble the high-contrast features formed in our experiment. Instead, we demonstrate the effect misalignments can have on aperture polarimetric measurements and polarimetric maps. We encourage exercising caution when interpreting polarimetric maps, particularly in cases where regions of high-contrast polarisation are found close to the photocentre. 

Finally, we would like to touch on the conclusion reached by \cite{bagnulo2021}, who equated the polarimetrically homogeneous coma of interstellar comet 2I/Borisov to it being a truly pristine comet, likely never having passed close to a star in its lifetime. As shown in this study, however, 67P also shows a homogeneous spread of polarisation throughout the coma despite orbiting the Sun every 6.45 years. Thus, we are interested in obtaining additional high resolution polarimetric data of comets of various types to further explore this topic.

%%%%%%%%%%%%%%%%%%%%%%%%%%%%%%%%%%%%%%%%%%%%%%%%%%

\section{Summary and Conclusions}

Using the ISIS instrument at the WHT and FORS2 at the VLT, we performed mutliwavelength ground-based imaging polarimetric observations of comet 67P in its 2015-16 apparition, coinciding with the Rosetta space mission, and its 2021 apparition. Our 2015-16 observations took place from four to seven months after perihelion, when the comet was at heliocentric distances r $\sim 1.9-2.5$ AU, while our 2021 observations occurred from six months to two weeks pre-perihelion. Overall, our observations cover a phase angle range of $4-50^{\circ}$. With these new observations, of unprecedentedly high S/N, we have performed aperture polarimetry and studied the phase angle dependence of polarisation, as well as plotted imaging and polarimetric maps. We summarise our findings as following:

\begin{enumerate}

\item Our new polarimetric measurements of 67P are comparable to previous studies, and we associate differences to the choice of aperture chosen and/or differences in data reduction processes. We confirm that the polarimetric phase angle dependence of 67P is typical for that of other JFCs and, according to our calculation of the best fit and the formal $1 \sigma$ uncertainties, is characterised by $\pmin = (-1.60 \pm 0.06) \%\ $ at $\amin = (9.1 \pm 0.1)^{\circ}$, and $\ainv = (20.2 \pm 0.1)^{\circ}$ with slope $h = (0.26 \pm 0.03)$.

\item We studied the difference between the pre- and post-perihelion polarimetric phase curves of 67P and found only a small difference between the two stages. This finding is only marginal evidence of the evolution in dust properties throughout the comet's orbital stages. Additional pre-perihelion measurements at small phase angles are necessary to reach a solid conclusion.

\item According to our imaging maps, 67P displayed an extended coma and tail throughout all observations. The isophotes plotted on these maps, along with the images enhanced through Laplace filtering and coma renormalisation, highlight areas of localised comet activity which can be tracked from epoch to epoch. Our observations reveal heightened activity concentrated in the northern region of the nucleus in early 2021, which later diminished as the comet approached perihelion, giving way to dominant activity in the southern hemisphere. Overall, this progression in localised cometary activity, as well as the geometry and orientation of the cometary structures detected in our maps, are comparable to previous studies of earlier apparitions of the comet.

\item Our polarimetric maps and aperture polarimetric measurements show that 67P is characterised by a homogeneous distribution of polarisation around the photocentre, with a lack of strong features or gradients in polarisation despite the detection of jets and fan-like features in the imaging maps. We attribute the homogeneous polarisation in the observed phase angle range to the prevalence of dust grains in the geometric optics regime, whose polarisation tend to converge to a single curve and, hence, value irrespective of particle size \citep{markkanen2018}. This provides an explanation for the absence of structures or gradients in the polarimetric maps. We are interested in investigating the possible presence of features in the polarimetric maps of cometary comae at larger phase angles, where differences may be found for particles of various sizes.

\end{enumerate}

In additional to these main findings, we investigated the consequences of deliberately introducing minor misalignments to the comet photocentre when combining the individual polarimetric frames, specifically in relation to the polarimetric maps and aperture polarimetric measurements. In our experiment, we found that even minor misalignments (values between --1 and 1 pixels) resulted in areas of high-contrast polarisation near the photocentre in the polarimetric maps, and created spurious, misleading features which could be mistaken for localised cometary activity. Despite these sharp gradients in polarisation, the aperture polarimetric measurements of these misaligned images show that the \pq\ values converge at larger aperture sizes. While we do not imply the inaccuracy of the polarimetric maps of studies that find variations in polarisation across the coma, we demonstrate that comparable values may be measured with aperture polarimetry even when polarimetric maps do not match.

\section*{Acknowledgements}
The observations presented in this work were made with the FORS2 instrument at the ESO Telescopes at the La Silla Paranal Observatory under program IDs 105.20LB.001, 105.20LB.002, 096.C-0821(A) and 096.C-0821(B), and with the ISIS instrument at the William Herschel Telescope (operated on the island of La Palma by the Isaac Newton Group), under programmes ITP2015-06. Research by Z. G. is funded by the UK Science and Technology Facilities Council (STFC).  Research by F.M. and O.M. has been partially supported by PID2021-123370OB-100/AEI/10.13039/501100011033/FEDER

%%%%%%%%%%%%%%%%%%%%%%%%%%%%%%%%%%%%%%%%%%%%%%%%%%
\section*{Data Availability}
All data are available at the ESO archive at {\tt archive.eso.org} and at the Astronomical Data Centre at {\tt http://casu.ast.cam.ac.uk/casuadc/}.



\bibliographystyle{mnras}
\bibliography{references} % if your bibtex file is called example.bib

%%%%%%%%%%%%%%%%%%%%%%%%%%%%%%%%%%%%%%%%%%%%%%%%%%

%%%%%%%%%%%%%%%%% APPENDICES %%%%%%%%%%%%%%%%%%%%%

%\appendix

% \section{Some extra material}

% If you want to present additional material which would interrupt the flow of the main paper,
% it can be placed in an Appendix which appears after the list of references.

%%%%%%%%%%%%%%%%%%%%%%%%%%%%%%%%%%%%%%%%%%%%%%%%%%

% Don't change these lines
\bsp	% typesetting comment
\label{lastpage}
\end{document}